\documentclass[prd,twocolumn,superscriptaddress,nofootinbib,showpacs]{revtex4}
\usepackage{graphicx}% Include figure files
\usepackage{dcolumn}% Align table columns on decimal point
\usepackage{bm}% bold math
\begin{document}
%\preprint{APS/123-QED}

\title{Anomalous CMB north-south asymmetry}

\author{Armando Bernui}
\email[]{bernui@das.inpe.br}
\affiliation{Instituto Nacional de Pesquisas Espaciais \\
             Divis\~{a}o de Astrof\'{\i}sica  \\
Av. dos Astronautas 1758, 12227-010 -- S\~ao Jos\'e dos Campos, SP, Brazil}

\date{\today}

\begin{abstract}
Several accurate analyses have revealed a statistically significant North-South 
ecliptic asymmetry in the large-angle correlations strength of the Cosmic Microwave 
Background (CMB) radiation temperature field data from the Wilkinson Microwave 
Anisotropy Probe (WMAP). 
This asymmetry is inconsistent with the statistical isotropy expected in the 
concordance cosmological model $\Lambda$CDM. 
It has been suggested that a possible cause-effect relationship exists between 
this large-angle anisotropy and the anomalous CMB quadrupole-octopole planes 
alignment. 
In turn, this later phenomenon (or both) would be a consequence of one or more of 
the following undesired effects in CMB data: a systematic error in the data 
processing or in the instrument characterization, residual foregrounds, and 
large-angle correlations induced by the incomplete sky CMB data (cut-sky masks are 
needed to reject Galactic foregrounds). 
Here, it is proved that the North-South asymmetry is unrelated to the 
quadrupole ($\ell=2$) and the octopole ($\ell=3$) properties because we find, 
at high confidence levels, such large-angle anisotropy in three- and five-year 
WMAP CMB maps containing only the multipole components $4 \leq \ell \leq 10$. 
The statistical significance depends on both, the CMB map analyzed as well as 
the cut-sky mask applied to exclude foregrounds. 
In general, we obtain that the significance level of the North-South asymmetry is 
less in five-year WMAP data with KQ75 ($\gtrsim 90\%$ CL) than it is in three-year 
data with Kp0 ($\gtrsim 96\%$ CL). 
For instance, in the WMAP ILC-five-year map with the KQ75 mask (a sky cut of 
28.4\%) this phenomenon is observed at 92.7\% CL, whereas for the WMAP 
ILC-three-year map with the Kp0 mask (a sky cut of 23.5\%) this phenomenon 
appears at 96.5\% CL. 
Moreover, it is also shown that this hemispherical asymmetry is unlikely due to 
systematics or foreground contaminants, because it is present in single-frequency, 
multifrequency, and cleaned ILC-type CMB maps. 
Additionally, robustness tests show that the statistical significance of the 
North-South asymmetry is affected by the use of the KQ85 and KQ75 masks in 
five-year single-frequency WMAP CMB maps, whereas it is insensitive with respect 
to application of the Kp2 and Kp0 masks in three-year single-frequency WMAP CMB 
maps. 
The confidence levels were obtained using sets of Monte Carlo CMB maps produced 
according to the $\Lambda$CDM model. 
\end{abstract}


\maketitle

%%%%%%%%%%%%%%%%%%%%%%%%%%%%%%%%%%%%%%%%%%%%%%%%%%%%%%%%%%%%%%%%%%%%%%%%%%%%%%%%
\section{Introduction} \label{Introduction}
%%%%%%%%%%%%%%%%%%%%%%%%%%%%%%%%%%%%%%%%%%%%%%%%%%%%%%%%%%%%%%%%%%%%%%%%%%%%%%%%

The five-year data release from the Wilkinson Microwave Anisotropy Probe 
(WMAP)~\cite{Hinshaw08,Gold08,Nolta08,Dunkley08,Komatsu08} have recently 
confirmed the validity of the concordance cosmological model $\Lambda$CDM, 
which successfully describes the contents and evolution of the Universe. 
These, almost full-sky, precise cosmological data of\/fer the unusual possibility 
to scrutinize the large-scale properties of the Universe, in particular those 
expected within this concordance model. 
One of these features concerns the set of Cosmic Microwave Background (CMB) 
radiation temperature f\/luctuations which is assumed to be a stochastic realization 
of a Gaussian random field, implying that their angular distribution on the celestial 
sphere is statistically isotropic at all angular scales. 
Examinations of the CMB maps from the first-year~\cite{Bennett03a,Bennett03b,Hinshaw03} 
and three-year WMAP data~\cite{Hinshaw06,Jarosik06,Spergel06} have revealed 
highly significant departures from statistical isotropy, at large angular scales. 
Evidences of anomalous large-angle anisotropies come from the power asymmetry 
of the CMB angular correlations between the northern and southern ecliptic 
hemispheres (hereafter NS-asymmetry), with indications of a preferred axis of 
maximum hemispherical asymmetry~\cite{Hansen04a,%
Hansen04b,Bielewicz04,Eriksen04a,Eriksen05,Land04,Land05b,Wiaux06,Copi06,Huterer06,%
Hansen06,PPG,Vielva06,Vielva07,Land07,Eriksen07,Copi05,BVWLF,Samal,BMRT,BMRT07,%
Bunn08,Monteserin,Lew} (for a dif\/ferent point of view see, e.g.,~\cite{Hajian03,%
Hajian04,Souradeep04,Souradeep05,Souradeep06,Hajian06}). 
Furthermore, other manifestations of CMB large-angle anomalies include the 
unlikely quadrupole-octopole planes alignment (referring both to the strong 
planarity of these multipoles as well as to the alignment between such 
planes~\cite{TOH,OTZH,Weeks04,Wiaux06,Copi06,Abramo06,PPG}).
Other unexpected results include deviations from Gaussianity~\cite{Chiang03,%
Vielva04,Copi04,Cruz05,Cruz07,McEwen06,BTV07,Martinez08}. 

Attempts to explain anomalous CMB phenomena have also considered globally 
axisymmetric space-times~\cite{Aurich05,Land05a,%
Jaffe06,Hipolito,Cresswell,Ghosh}, motivated by the fact that CMB data seem 
to indicate a preferred direction in the space. 
Other studies, instead, have investigated physical mechanisms that break statistical 
isotropy, or during the epoch of inf\/lation~\cite{Gordon05,Contaldi07,Pullen,Ackerman,%
Dutta,Dvorkin} or during the decoupling era~\cite{Demianski,Pereira,Campanelli2,%
Morales,BH,Kahniashvili1,Kahniashvili2}, in order to account for two (or more) 
large-angle CMB anomalies. 
The motivation for searching a scenario that correlates two CMB anomalies 
is to comprehend whether these phenomena have dif\/ferent causes or instead they 
are manifestations of a unique ef\/fect. 
For this, a number of studies have searched for a possible relationship 
between the NS-asymmetry and the CMB quadrupole-octopole alignment phenomena (see, 
e.g.,~\cite{Wiaux06,Rakic07,Land07,BH}). 

Although the origin of these CMB anomalies remains unknown, possible sources 
include CMB residual foregrounds~\cite{Eriksen04b,WW,%
OT,Cruz06,ASW,Lopez07,Chiang07}, and systematic errors~\cite{Copi05,Copi06,%
Helling06,Bunn07}. 
In particular, there are indications that unremoved contaminants~\cite{Naselsky07} 
or omitted systematics~\cite{Schwarz04} could be inf\/luencing the reconstruction 
of the CMB quadrupole and octopole. 
Moreover, the WMAP cut-sky masks used to remove residual foregrounds probably 
induce spurious large-angle correlations, producing forged low-order CMB 
multipoles~\cite{Bielewicz05}. 
Thereby ef\/forts are being done by the WMAP science team to improve the 
data processing by minimizing the ef\/fects caused by foregrounds and systematic 
errors~\cite{Jarosik06,Gold08}. 

Here we show that the NS-asymmetry phenomenon is not related to the CMB's 
quadrupole or octopole components. 
To obtain this result we perform a two-point angular correlations analysis 
in an array of large-angle spherical caps that scan the CMB sky, where the  
WMAP maps investigated have their corresponding quadrupole and octopole 
components removed. 
The outline of this paper is the following. 
In Sec.~\ref{method}, we present the geometrical-statistical method, termed 
{\em sigma map}, that led us to investigate the directional angular correlations 
strength in a set of WMAP CMB maps. 
In Sec.~\ref{CMB analyses} we first discuss the maps to be analyzed, after 
that we apply our method to these CMB data, where the statistical confidence is 
evaluated according to Monte Carlo maps produced according to the $\Lambda$CDM model. 
Finally, in Sec.~\ref{conclusions}, we formulate our conclusions and final remarks.

%%%%%%%%%%%%%%%%%%%%%%%%%%%%%%%%%%%%%%%%%%%%%%%%%%%%%%%%%%%%%%%%%%%%%%%%%%%%%%%%%%%%
\section{The 2PACF and the Sigma-Map method}   \label{method}
%%%%%%%%%%%%%%%%%%%%%%%%%%%%%%%%%%%%%%%%%%%%%%%%%%%%%%%%%%%%%%%%%%%%%%%%%%%%%%%%%%%%

Our method to investigate the large-scale angular correlations in CMB temperature 
f\/luctuations maps consists in the computation of the 2-point angular correlation 
function (2PACF)~\cite{Padmanabhan} in a set of spherical caps covering the 
celestial sphere. 

Let $\Omega_{\gamma_0}^J \equiv \Omega(\theta_J,\phi_J;\gamma_0) 
\subset {\cal S}^2$ be a spherical cap region on the celestial sphere, 
of $\gamma_0$ degrees of aperture, with vertex at the $J$-th pixel, 
$J = 1, \ldots, N_{\mbox{\footnotesize caps}}$, where $(\theta_J,\phi_J)$ 
are the angular coordinates of the $J$-th pixel's center. 
Both, the number of spherical caps $N_{\mbox{\footnotesize caps}}$ and the 
coordinates of their centers $(\theta_J,\phi_J)$ are def\/ined using the HEALPix 
pixelization scheme~\cite{Gorski}. 
The union of the $N_{\mbox{\footnotesize caps}}$ spherical caps covers completely 
the celestial sphere ${\cal S}^2$. 

Given a pixelized CMB map, the 2PACF of the temperature f\/luctuations 
$\delta T$ corresponding to the pixels located in the spherical cap 
$\Omega_{\gamma_0}^J$ is def\/ined by~\cite{Padmanabhan} 
\begin{equation} \label{2PACF}
\mbox{\rm C}(\gamma)^J \equiv 
\langle\, \delta T(\theta_i,\phi_i) \delta T(\theta_{i'},\phi_{i'}) \,\rangle \, ,
\end{equation}
where 
$\cos\gamma = \cos\theta_i \cos\theta_{i'} 
+ \sin\theta_i \sin\theta_{i'} \cos(\phi_i\!-\!\phi_{i'})$, 
and $\gamma \in (0,2\gamma_0]$ is the angular distance between the $i$-th and 
the $i'$-th pixels centers. The average $\langle \,\, \rangle$ in the above
equation is done over all the products 
$\delta T(\theta_i,\phi_i) \delta T(\theta_{i'},\phi_{i'})$ such that 
$\gamma_k \equiv \gamma \in ((k-1)\delta,\, k\delta]$, for 
$k = 1,..., N_{\mbox{\footnotesize bins}}$, where 
$\delta \equiv 2\gamma_0 / N_{\mbox{\footnotesize bins}}$
is the bin-width. 
We denote by $\mbox{\rm C}_k^J \equiv \mbox{\rm C}(\gamma_k)^J$ the value of the 
2PACF for the angular distances $\gamma_k \in ((k-1)\delta,\, k\delta]$. 
%
%%%%%%%%%%%%%%%%%%%%%%%%%%%%%%%%  sigma-map  %%%%%%%%%%%%%%%%%%%%%%%%%%%%%%%
%
Def\/ine now the scalar function 
$\sigma: \Omega_{\gamma_0}^J \subset {\cal S}^2 \mapsto {\Re}^{+}$, 
for $J = 1, \ldots, N_{\mbox{\footnotesize caps}}$, which assigns to the 
$J$-cap, centered at $(\theta_J,\phi_J)$, a real positive number 
$\sigma_J \equiv \sigma(\theta_J,\phi_J) \in \Re^+$. 
The most natural way of def\/ining a measure $\sigma$ is through the 
variance of the $\mbox{\rm C}_k^J$ function~\cite{BMRT}, 
\begin{equation} \label{sigma}
\sigma^2_J  \equiv \frac{1}{N_{\mbox{\footnotesize bins}}}
\sum_{k=1}^{N_{\mbox{\footnotesize bins}}} (\mbox{\rm C}_k^J)^2 \,\, . 
\end{equation}
To obtain a quantitative measure of the angular correlations in a CMB map, we 
cover the celestial sphere with $N_{\mbox{\footnotesize caps}}$ spherical caps, 
and calculate the set of sigma values 
$\{ \sigma_J, \, J=1,...,N_{\mbox{\footnotesize caps}} \}$ using Eq.~(\ref{sigma}). 
Associating the $J$-th sigma value $\sigma_J$ to the $J$-th pixel, for 
$J=1, \ldots, N_{\mbox{\footnotesize caps}}$,  one f\/ills the celestial 
sphere with positive real numbers, and according to a linear scale (where 
$\sigma^{\mbox{\footnotesize minimum}} \rightarrow blue$,  
$\sigma^{\mbox{\footnotesize maximum}} \rightarrow red$), one converts this 
numbered map into a colored map: this is the sigma map. 
F\/inally, we f\/ind the multipole components of a sigma map by calculating 
its angular power spectrum. 
In fact, given a sigma map one can expand $\sigma = \sigma(\theta,\phi)$ in 
spherical harmonics: 
$\sigma(\theta,\phi) = \sum_{\ell,\, m} A_{\ell\, m} Y_{\ell\, m}(\theta,\phi)$. 
Then, the set of values $\{ \mbox{\sc S}_{\ell},\, \ell=1,2,... \}$, where 
$\mbox{\sc S}_{\ell} \equiv (1 / (2\ell+1)) 
\sum_{m={\mbox{\small -}}\ell}^{\ell} \, |A_{\ell\, m}|^2$, 
give the angular power spectrum of the sigma map.

A power spectrum $\mbox{\sc S}_{\ell}^{_{\mbox{\footnotesize WMAP}}}$ of a 
sigma map computed from a given WMAP map, provides quantitative information 
about large-angle anisotropy features of such a CMB map when compared with the 
mean of sigma-map power spectra obtained from Monte Carlo CMB maps produced under 
the statistical isotropy hypothesis. 
As we shall see, the sigma-map analysis is a suitable tool to reveal large-angle 
anisotropies in CMB maps, like the NS-asymmetry.

%%%%%%%%%$%%%%%%%%%%%%%%%%%%%%%%%%%%%%%%%%%%%%%%%%%%%%%%%%%%%%%%%%%%%%%%%%%%%%%%%% 
\section{CMB analyses with the sigma map}  \label{CMB analyses}
%%%%%%%%$%%%%%%%%%%%%%%%%%%%%%%%%%%%%%%%%%%%%%%%%%%%%%%%%%%%%%%%%%%%%%%%%%%%%%%%%% 

In this section, we apply the sigma-map method to study the large-angle correlations 
in a set of single- and multifrequency CMB temperature f\/luctuations maps, all 
coming from the three-year and five-year WMAP data.

%%%%%%%%%%%%%%%%%%%%%%%%%%%%%%%%%%%%%%%%%%%%%%%%%%%%%%%%%%%%%%%%%%%%%%%%%%%%%%%%%%% 
\subsection{The WMAP data}  \label{WMAP data}
%%%%%%%%%%%%%%%%%%%%%%%%%%%%%%%%%%%%%%%%%%%%%%%%%%%%%%%%%%%%%%%%%%%%%%%%%%%%%%%%%%% 

The WMAP science team made substantial ef\/forts to improve the data products by 
minimizing the contaminating ef\/fects caused by dif\/fuse Galactic foregrounds, 
astrophysical point sources, artifacts from the instruments and measurement process, 
and systematic errors~\cite{Jarosik06,Gold08}. 
As a result, it has been released as the single-frequency CMB foreground-reduced Q-, 
V-, and W-band maps corresponding to the less contaminated frequencies centered at 41, 
61, and 94 GHz, respectively (hereafter the Q-5yr, V-5yr, and W-5yr maps, respectively), 
and an Internal Linear Combination (ILC-5yr) full-sky CMB map~\cite{Hinshaw06}. 
Moreover, the eight foreground-reduced dif\/ferencing assemblies (two in the Q-band, 
two in the V-band, and four in the W-band) can be combined using inverse-noise-variance 
weights to produce multifrequency CMB maps, like the QVW and the VW coadded maps, 
hereafter termed QVW-5yr and VW-5yr maps~\cite{Hinshaw03}. 
The corresponding three-year maps are the Q-3yr, V-3yr, W-3yr, ILC-3yr, QVW-3yr, 
and VW-3yr CMB maps~\cite{Hinshaw06}. 
In addition to these maps, other teams have also produced full-sky cleaned 
CMB maps using the unprocessed three-year and five-year data through dif\/ferent 
foreground cleaning tools, that is, the de Oliveira-Costa-Tegmark~\cite{OT}, 
the Park-Park-Gott~\cite{PPG}, and the Kim-Naselsky-Christensen CMB maps~\cite{KNC}, 
hereafter termed the OT-3yr, PPG-3yr, and KNC-5yr, respectively. 

In order to validate our sigma-map results we shall estimate their sensitivity 
to the fol\/lowing ef\/fects, which are possible sources of large-angle anomalies 
in WMAP data: 
(i)   residual foregrounds, 
(ii)  masking applications, 
(iii) systematic errors (coming from, e.g., instrument noise, 
foreground-minimization process, time-ordered data analysis, etc.). 
Regarding the issue (i), we know that the Galactic foregrounds are frequency 
dependent~\cite{Bennett03b}. 
For this reason the study of the individual frequency CMB maps, that is the 
foreground-reduced Q, V, and W three- and five-year maps are essential to 
detect the possible contribution of Galactic contaminants in our results. 
Moreover, following the WMAP science team recommendation to reject foregrounds 
contamination in temperature analyses in five-year (three-year) data we use 
the Galactic mask KQ85 (Kp2), which cuts 18.3\% (15.4\%) of the sky 
data~\cite{Gold08,Nolta08}. 
Regarding issue (ii), we shall test the sensitivity of our results due to 
the masking application by comparing the ef\/fects due to the more severe %conservative 
five-year KQ75 (three-year Kp0) mask, which cuts 28.4\% (23.5\%) of the sky data,
with respect to the KQ85 (Kp2) mask. 
Regarding issue (iii), we know that (large-angle) systematic ef\/fects leave 
(large-angle) imprints in the CMB maps~\cite{Jarosik06,Hinshaw08,Gold08,Dunkley08}. 
Because we are investigating a large-scale anisotropy, that is the NS-asymmetry 
phenomenon, we intend to avoid possible large-angle systematic ef\/fects. 
According to recent reports the quadrupole and the octopole CMB components could 
be contaminated with systematics, e.g., in the time-ordered data 
analyses~\cite{Schwarz04,Copi05,Copi06}, or due to the masking 
procedures~\cite{Bielewicz05}. 
Furthermore, systematics could also appear in the foreground-decontamination 
processes~\cite{Lopez07,Chiang07,Helling06,Bunn07,Naselsky07}. 
For these reasons, we study the large-scale properties of the CMB temperature field 
in WMAP data containing just the multipoles $4 \leq \ell \leq 10$, that is, removing 
the quadrupole ($\ell=2$) and the octopole ($\ell=3$) CMB components. 
Moreover, we investigate all the publicly available cleaned maps just to exclude 
the possibility that a particular foreground-minimization process is introducing 
systematic ef\/fects in the data. 

Our aim is to investigate the large-scale angular distribution of the CMB 
temperature f\/luctuations. 
Given a CMB map, we shall concentrate on the submap containing the set of 
multipole components $4 \leq \ell \leq 10$. 
To produce this submap, containing the CMB temperature f\/luctuations information 
at these angular scales, we first use the {\sc anafast} code~\cite{Gorski} to obtain, 
after applying a cut-sky mask, the $\{ a_{\ell m} \}$ values corresponding to 
these multipoles. 
Then, using the {\sc synfast} code~\cite{Gorski}, we use this set $\{ a_{\ell m} \}$ 
to make up the CMB map. 
We emphasize that, all the CMB maps here investigated using the sigma-map analysis 
only contain the multipole components $4 \leq \ell \leq 10$, and we adopt the 
following notation: 
for instance, the ILC-5yr-KQ85 map is obtained from the ILC-5yr CMB map by 
selecting, after applying the KQ85 mask, those multipoles in the range 
$4 \leq \ell \leq 10$. 

We generate one set of $1\,000$ Monte Carlo CMB maps (MC) which correspond to 
random realizations seeded by the $\Lambda$CDM angular power spectra~\cite{Nolta08}, 
with $N_{\mbox{\footnotesize\rm side}}=512$, and $\ell_{max}=1\,024$. 
From this MC set we produced two sets of $1\,000$ MC each for the multipole range 
$4 \leq \ell \leq 10$ after using the KQ85 and KQ75 masks, respectively. 
The statistical significance analyses come from the comparison between the angular 
power spectra of the sigma maps obtained from the WMAP data (hereafter sigma-maps 
WMAP) and the corresponding angular power spectra calculated from the sigma maps 
computed from the MC CMB maps (hereafter sigma-maps MC).

%%%%%%%%%%%%%%%%%%%%%%%%%%%%%%%%%%%%%%%%%%%%%%%%%%%%%%%%%%%%%%%%%%%%%%%%%%%%%%%% 
\subsection{Data analyses with the KQ85 and KQ75 masks}  \label{results1}
%%%%%%%%%%%%%%%%%%%%%%%%%%%%%%%%%%%%%%%%%%%%%%%%%%%%%%%%%%%%%%%%%%%%%%%%%%%%%%%% 

Our objective here is to produce the sigma-maps WMAP for each of the five-year 
(three-year) WMAP maps mentioned above, containing the multipoles 
$4 \leq \ell \leq 10$ obtained after using the KQ85 (Kp2) cut-sky mask. 
After the calculation of the sigma-maps WMAP we compute their corresponding 
angular power spectra $\{ \mbox{\sc S}_{\ell}, \ell=1,...,5 \}$ in order to 
quantify, by comparison with the corresponding set of sigma-maps MC, if they 
are consistent, or not, with statistically isotropic $\Lambda$CDM CMB maps. 

\begin{figure} 
\vspace{-1cm}
\includegraphics[width=15cm,height=24cm]{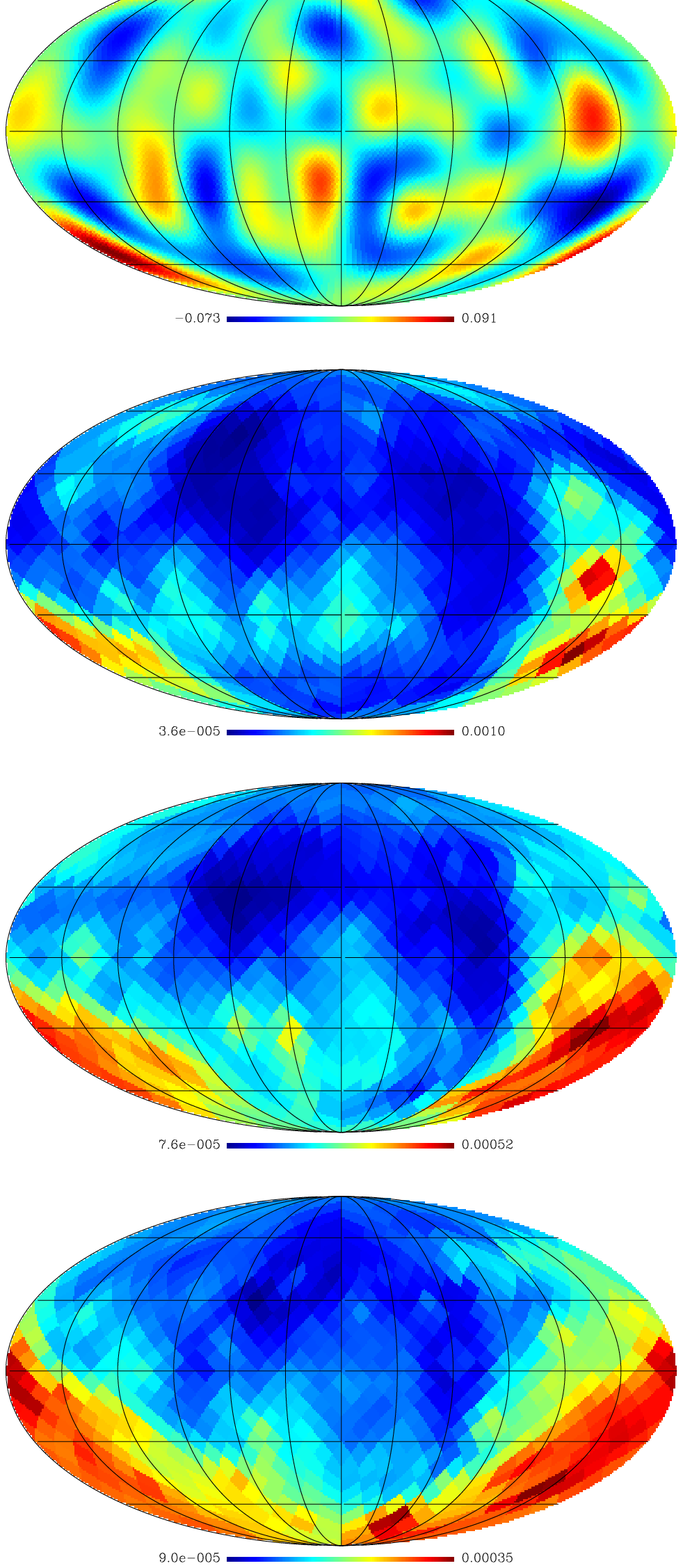}

\vspace{-3cm}
\caption{\label{fig1} 
The top panel shows the CMB map ILC-5yr-KQ85-$\ell_{4-10}$.
The second, third, and fourth panels are the sigma-maps calculated from this 
CMB map considering spherical caps with apertures 
$\gamma_0 = 30^{\circ}, 45^{\circ}, 60^{\circ}$, respectively. 
The NS-asymmetry is clearly seen in the uneven hemispherical distribution 
of the angular correlations strength, depicted as red and blue pixels (large 
and small sigma-values, respectively). 
All the maps are plotted in Galactic coordinates.} 
\end{figure}

\begin{figure} 
\vspace{-1cm}
\includegraphics[width=15cm,height=24cm]{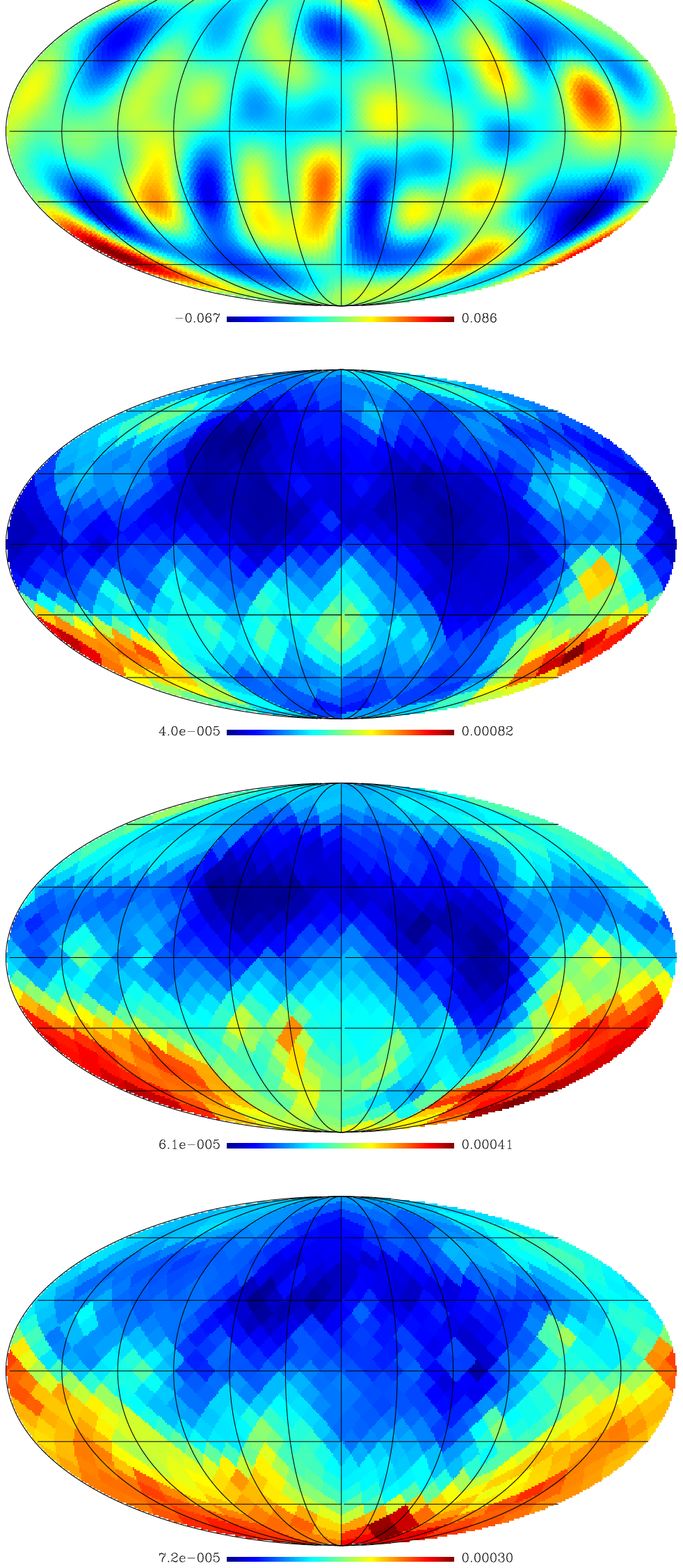}

\vspace{-3cm}
\caption{\label{fig2} 
The top panel shows the CMB map ILC-5yr-KQ75-$\ell_{4-10}$.
The second, third, and fourth panels are the sigma-maps calculated from 
this CMB map considering spherical caps with apertures 
$\gamma_0 = 30^{\circ}, 45^{\circ}, 60^{\circ}$, respectively. 
Again, the NS-asymmetry is clearly seen in the uneven hemispherical distribution 
of the angular correlations strength, depicted as red and blue pixels (large 
and small sigma-values, respectively). 
All the maps are plotted in Galactic coordinates.
}  
\end{figure}

To illustrate our results we show in Fig.~\ref{fig1} (Fig.~\ref{fig2}), 
from top to bottom, the ILC-5yr-KQ85 (ILC-5yr-KQ75) CMB map, and its 
corresponding sigma-maps WMAP, computed for three dif\/ferent $\gamma_0$ values, 
that is, $\gamma_0 = 30^{\circ}, 45^{\circ}, 60^{\circ}$. 
As we observe, these sigma-maps WMAP clearly exhibit a dipolar red-blue region
strongly indicative of a large-angle asymmetry in the distribution of the 
angular correlations, the so-called NS-asymmetry phenomenon 
(in all sky maps we have used Galactic coordinates where the equator is defined 
by the latitude $b=90^\circ$, with the Galactic center $(l, b) = (0,90^\circ)$ 
in the middle of the figure, and the longitude $l$ increases to the left). 

To quantify this NS-asymmetry in the sigma-maps-WMAP-KQ85 and 
sigma-maps-WMAP-5yr-KQ75, we calculate their angular power spectra, 
and then evaluate their statistical significance by comparison with the 
corresponding sigma-maps MC's angular power spectra. 
Our results appear in the top-panel plots of Figs.~\ref{fig3} and~\ref{fig4} 
corresponding to the WMAP-KQ85 and WMAP-5yr-KQ75 cases, 
respectively, where solid lines represent the mean values obtained from 
sigma-maps MC and dashed lines represent the 95\% CL from these data. 
In each case the MC data contain the multipoles $4 \leq \ell \leq 10$ obtained 
after applying the corresponding KQ85 or KQ75 cut-sky mask. 
Notice that in Figs.~\ref{fig3} and~\ref{fig4} the symbols representing data from 
three-year sigma-maps WMAP are plotted (in red) close to the vertical axis, while 
the symbols representing data from the five-year sigma-maps WMAP are plotted 
(in blue) slightly shifted to the right. 
This is done in order to observe the dif\/ferences outcoming from distinct data 
releases. 
For illustration, in the bottom panels of Figs.~\ref{fig3} and~\ref{fig4} we 
plotted the dipole components of the sigma-map-ILC-5yr KQ85 and 
sigma-map-ILC-5yr KQ75, respectively, where we observe that both dipoles point 
nearly in the same direction $(l,b) \simeq (180^{\circ},130^{\circ})$, 
which means that this result is stable with respect to the choice of the KQ85 
and KQ75 Galactic masks. 

The sigma-maps WMAP angular power spectra shown at the top panels of Figs.~\ref{fig3} 
and~\ref{fig4}, were obtained using spherical caps of aperture 
$\gamma_0 = 45^{\circ}$, $N_{\mbox{\footnotesize bins}}=45$, and 
$N_{\mbox{\footnotesize caps}}=768$. 
For calculating the sigma maps we used CMB maps with pixelization parameter 
$N_{\mbox{\footnotesize side}}=32$. 
Concerning the robustness of our results, we studied the ef\/fects of changing  
various parameters employed in the calculation of the sigma maps. 
Our tests included the computation of the sigma maps using spherical caps of 
$\gamma_0 = 30^{\circ},\, 45^{\circ},\, 60^{\circ}$ of aperture, also 
considering dif\/ferent $N_{\mbox{\footnotesize bins}}=45,\, 90,\, 120$ values, 
and $N_{\mbox{\footnotesize caps}}=768, \,\, 3\,072$. 
Additionally, we also studied the inf\/luence of the angular resolution of the 
CMB maps by computing the sigma-maps WMAP with dif\/ferent pixelizations parameter, 
namely $N_{\mbox{\footnotesize side}}=16,\, 32$.
We found that the sigma-maps WMAP as well as their corresponding angular power 
spectra calculated in these cases are completely similar to those results exhibit 
in Figs.~\ref{fig3} and~\ref{fig4}.

Examining data from the sigma-maps-WMAP KQ85, shown in the top panel of 
Fig.~\ref{fig3}, we observe that the Q-3yr (bullet), V-3yr (asterisk), 
and W-3yr (star) data do not show significative dif\/ferences between them, 
but the Q-5yr (bullet), V-5yr (asterisk), and W-5yr (star) data exhibit very 
dif\/ferent dipole values. 
In fact, the confidence level values for the dipole values 
$\mbox{\sc S}_{1}^{_{\mbox{\footnotesize Q-3yr}}}$, 
$\mbox{\sc S}_{1}^{_{\mbox{\footnotesize V-3yr}}}$, 
$\mbox{\sc S}_{1}^{_{\mbox{\footnotesize W-3yr}}}$ are 95.1\%, 95.7\%, and 95.9\% CL, 
respectively, whereas for % media = 95.5667 \pm 0.416334
$\mbox{\sc S}_{1}^{_{\mbox{\footnotesize Q-5yr}}}$, 
$\mbox{\sc S}_{1}^{_{\mbox{\footnotesize V-5yr}}}$, 
$\mbox{\sc S}_{1}^{_{\mbox{\footnotesize W-5yr}}}$ are 85.9\%, 95.8\%, and 99.5\% CL, 
respectively.             % media = 93.7333 \pm 7.03160
These dif\/ferences in WMAP-KQ85 five-year data suggest the presence 
of residual foregrounds in Q-5yr, V-5yr, and W-5yr maps. 
On the other hand, the Q-3yr, V-3yr, and W-3yr maps appear to be foreground 
cleaned enough. 
Nevertheless, the quantitative analysis shown at the top panel of Fig.~\ref{fig3} 
confirms the NS-asymmetry phenomenon, at 95\% $-$ 99\% CL, for all (except one of) 
the WMAP-KQ85 CMB maps analyzed. 
The exception is due to the Q-5yr-KQ85 map. 
We emphasize that these results are independent of the CMB quadrupole ($\ell=2$) 
and octopole ($\ell=3$) components because they were completely removed 
before we began our analyses. 

Regarding the quantitative sigma-map analysis for the WMAP-KQ75 
CMB maps, top panel of Fig.~\ref{fig4}, we verify that the NS-asymmetry 
phenomenon is present, at more than 98\% CL, in all the single, multifrequency, 
and ILC-type WMAP-3yr-Kp0 maps and also in the KNC-5yr-KQ75 foreground cleaned 
map. 
In all the other WMAP-5yr-KQ75 maps the NS-asymmetry phenomenon appears 
with a lower statistical significance. 
It is worth noting, again, that the dipole values $\mbox{\sc S}_{1}$ from 
single frequency Q-, V-, W-5yr-KQ75 maps are disperse, on the 
contrary, again, the dipole values $\mbox{\sc S}_{1}$ from single frequency 
Q-, V-, W-3yr-Kp0 maps are clustered, a distinct behavior that suggests the 
presence of residual foregrounds, deserving further analysis. 

The possibility that foregrounds were not completely removed with the application 
of the KQ85 mask motivate further analysis of the WMAP data through the use of 
the more severe cut-sky KQ75 mask. 
Analyzing our results from the sigma-maps-WMAP-5yr KQ75 and 
sigma-maps-WMAP-3yr Kp0, shown in the top 
panel of Fig.~\ref{fig4}, we observe that the conf\/idence level for the dipole values 
$\mbox{\sc S}_{1}^{_{\mbox{\footnotesize Q-3yr}}}$, 
$\mbox{\sc S}_{1}^{_{\mbox{\footnotesize V-3yr}}}$, 
$\mbox{\sc S}_{1}^{_{\mbox{\footnotesize W-3yr}}}$ are 95.6\%, 96.3\%, and 96.7\% CL, 
respectively, whereas for % media = 96.2000 \pm 0.556776
$\mbox{\sc S}_{1}^{_{\mbox{\footnotesize Q-5yr}}}$, 
$\mbox{\sc S}_{1}^{_{\mbox{\footnotesize V-5yr}}}$, 
$\mbox{\sc S}_{1}^{_{\mbox{\footnotesize W-5yr}}}$ are 89.2\%, 90.9\%, and 93.7\% CL, 
respectively.             % media = 91.2667 \pm 2.27230
We observe that the application of the more severe KQ75 cut-sky mask produces a 
lower dispersion of the confidence levels for these dipoles (mean $=91.3 \pm 2.3$\% CL) 
as compared with the KQ85 mask case (mean $=93.7 \pm 7.0$\% CL), 
but it is still large as compared with the WMAP-3yr case. 
In fact, the dipole terms corresponding to the single-frequency Q-3yr, V-3yr, 
and W-3yr WMAP maps have their mean and standard deviation: $96.2 \pm 0.6$\% CL 
and $95.6 \pm 0.4$\% CL for the Kp0 and Kp2 masks, respectively. 
This later result reveals two important consequences regarding the single-frequency 
WMAP-3yr maps: first, these CMB maps appear suf\/ficiently foreground cleaned, 
and second, the NS-asymmetry phenomenon exhibited in three-year WMAP data is 
robust with respect to the Kp0 and Kp2 masks. 
From other side, the masking-dependent and disperse results achieved in the 
analysis of the single-frequency Q-5yr, V-5yr, and W-5yr maps could indicate 
possible unremoved foregrounds, regardless of the severe KQ75 mask applied. 

Additionally, a close inspection of the data plotted in the top panel of 
Fig.~\ref{fig4} reveals another interesting feature concerning the WMAP-ILC maps 
and the applied masks. 
We notice that the statistical significance for the dipole value of the 
sigma-map-ILC-5yr KQ75 is 92.7\% CL, 
whereas for the dipole value of the sigma-map-ILC-5yr Kp0 is 96.5\% CL, which 
shows a $\sim 4 \%$ dif\/ference. 
From other side, one verifies that the ILC-5yr-KQ75 and ILC-3yr-Kp0 CMB maps 
are fully similar as revealed by the Pearson's correlation coef\/ficient: 
0.985180 and 0.985466, for the CMB data outside the KQ75 and Kp0 masks, 
respectively, which are only $\sim 0.03 \%$ dif\/ferent. 
The question is, why highly correlated CMB maps give such dissimilar (sigma-map) 
results? 
The answer involves a twofold aspect of a given mask: the amount of the sky cutted 
by the mask and the location of the sky-patch cutted. 
These important issues, related to the robustness of our results, motivate a 
masking-ef\/fect analysis on the angular power spectra of the sigma-maps WMAP.

%%%%%%%%%%%%%%%%%%%%%%%%%%%%%%%%%%%%%%%%%%%%%%%%%%%%%%%%%%%%%%%%%%%%%%%%%%%%%%%% 
\subsection{Masking-ef\/fect analysis: KQ75 versus Kp0}  \label{results2}
%%%%%%%%%%%%%%%%%%%%%%%%%%%%%%%%%%%%%%%%%%%%%%%%%%%%%%%%%%%%%%%%%%%%%%%%%%%%%%%% 

Observing the data plotted in the top panel of Fig.~\ref{fig4}, one realizes 
that the spectra from sigma-map-3yr-Kp0 data have consistently 
larger values than the spectra from sigma-map-5yr KQ75. 
This fact could be indicating a systematic ef\/fect due to the dif\/ferent masks, 
KQ75 and Kp0, employed in these calculations. 
Clearly, the use of these two masks involves two ef\/fects: 
the dif\/ferent amount of cutted sky, of $\sim 5\%$ between the KQ75 and Kp0 
masks, and the sky location of the cutted patch. 
A consequence of the first ef\/fect is that dif\/ferent sky regions with CMB data 
are being considered to make up the WMAP-KQ75 and WMAP-Kp0 maps, which 
implies distinct contributions into their corresponding sigma maps; 
a consequence of the second ef\/fect is that depending on the location of the 
cutted region, the calculation of the 2PACF for the sigma map could be 
strengthened or weakened. 
In both cases, we are introducing (positive or negative) spurious correlations 
in the sigma maps whose inf\/luence on our results need to be examined. 
The study of the masking ef\/fects on the sigma maps is connected with the second 
issue mentioned in~\ref{WMAP data}. 

For this, we perform a comparative sigma-map analysis of the four CMB maps: 
ILC-3yr-KQ75, ILC-3yr-Kp0, ILC-5yr-KQ75, and ILC-5yr-Kp0. 
In other words, we shall test the sensitivity of the ILC-3yr and ILC-5yr CMB 
maps to the ef\/fect of interchanging the use of KQ75 and Kp0 masks to obtain 
the multipoles components $4 \leq \ell \leq 10$. 
Our results are shown in Fig.~\ref{fig5}.
In the top panel we plot the sigma-map spectra of these four CMB maps, that is, 
$\{ \mbox{\sc S}_{\ell}^{_{\mbox{\footnotesize ILC-3yr-KQ75}}} \}$,
$\{ \mbox{\sc S}_{\ell}^{_{\mbox{\footnotesize ILC-3yr-Kp0}}} \}$,
$\{ \mbox{\sc S}_{\ell}^{_{\mbox{\footnotesize ILC-5yr-KQ75}}} \}$, and
$\{ \mbox{\sc S}_{\ell}^{_{\mbox{\footnotesize ILC-5yr-Kp0}}} \}$, 
represented as circle, filled-circle, square, and filled-square symbols, 
respectively. 
Our results show 
\begin{eqnarray}
&\mbox{\sc S}_{\ell}^{_{\mbox{\footnotesize ILC-3yr-KQ75}}} \simeq 
\mbox{\sc S}_{\ell}^{_{\mbox{\footnotesize ILC-5yr-KQ75}}}& \, , \nonumber \\
&\mbox{\sc S}_{\ell}^{_{\mbox{\footnotesize ILC-5yr-Kp0}}} \simeq 
 \mbox{\sc S}_{\ell}^{_{\mbox{\footnotesize ILC-3yr-Kp0}}}& \, , \nonumber
\end{eqnarray}
proving that the main reason (if not the unique one) for the dif\/ferent 
spectra values corresponding to the three- and five-year sigma-maps ILC 
is the distinct cut-sky area defined by the KQ75 and Kp0 masks. 
Additionally, we find it interesting to know the contrasting ef\/fect produced 
by these two masks in the sigma-map-ILC-5yr KQ75 and sigma-map-ILC-5yr Kp0. 
For this we present at the bottom panel of Fig.~\ref{fig5} the dif\/ference map: 
sigma-map-ILC-5yr Kp0 minus sigma-map-ILC-5yr KQ75, which clearly reveals through 
its red-intense spots that the application of dif\/ferent masks has a significant 
impact on the detection of the NS-asymmetry phenomenon (the corresponding 
dif\/ference map for ILC-3yr case is fully similar). 
Observing this dif\/ference map we understand why the use of the KQ75 mask 
decreases the statistical significance of the NS-asymmetry phenomenon: 
this mask cuts a sky-patch that turns out to be critical for the computation 
of the large-scale 2PACF in the sigma-maps. 

To il\/lustrate the ful\/l-sky case, that is a sigma-map analysis of the 
foreground-cleaned ILC-5yr CMB map with multipoles $4 \le \ell \le 10$ obtained 
without using a mask, we show in Fig.~\ref{fig6} the sigma-map-ILC-5yr (top panel) 
and its corresponding dipole term (bottom panel). 
The statistical significance of this sigma-map-ILC 5yr, as compared with 1\,000 
sigma-maps MC obtained from an equal number of full-sky $\Lambda$CDM 
MC-$\ell_{4-10}$ CMB maps, is 95.8\% CL.

%%%%%%%%%%%%%%%%%%%%%%%%%%%%%%%%%%%%%%%%%%%%%%%%%%%%%%%%%%%%%%%%%%%%%%%%%%%%%%%% 
\section{Conclusions} \label{conclusions}
%%%%%%%%%%%%%%%%%%%%%%%%%%%%%%%%%%%%%%%%%%%%%%%%%%%%%%%%%%%%%%%%%%%%%%%%%%%%%%%% 

We investigated a set of foregrounds-reduced and systematics-cleaned 
single-frequency, multifrequency, and ILC-type CMB maps from three- and five-year 
WMAP data. 
We performed the sigma-maps analyses of these CMB maps at large-angles corresponding 
to the multipoles range $4 \leq \ell \leq 10$, 
and for dif\/ferent three- and five-year WMAP cut-sky masks, that is Kp2, Kp0, 
KQ85, and KQ75. 

\begin{figure} 
\includegraphics[width=15cm,height=24cm]{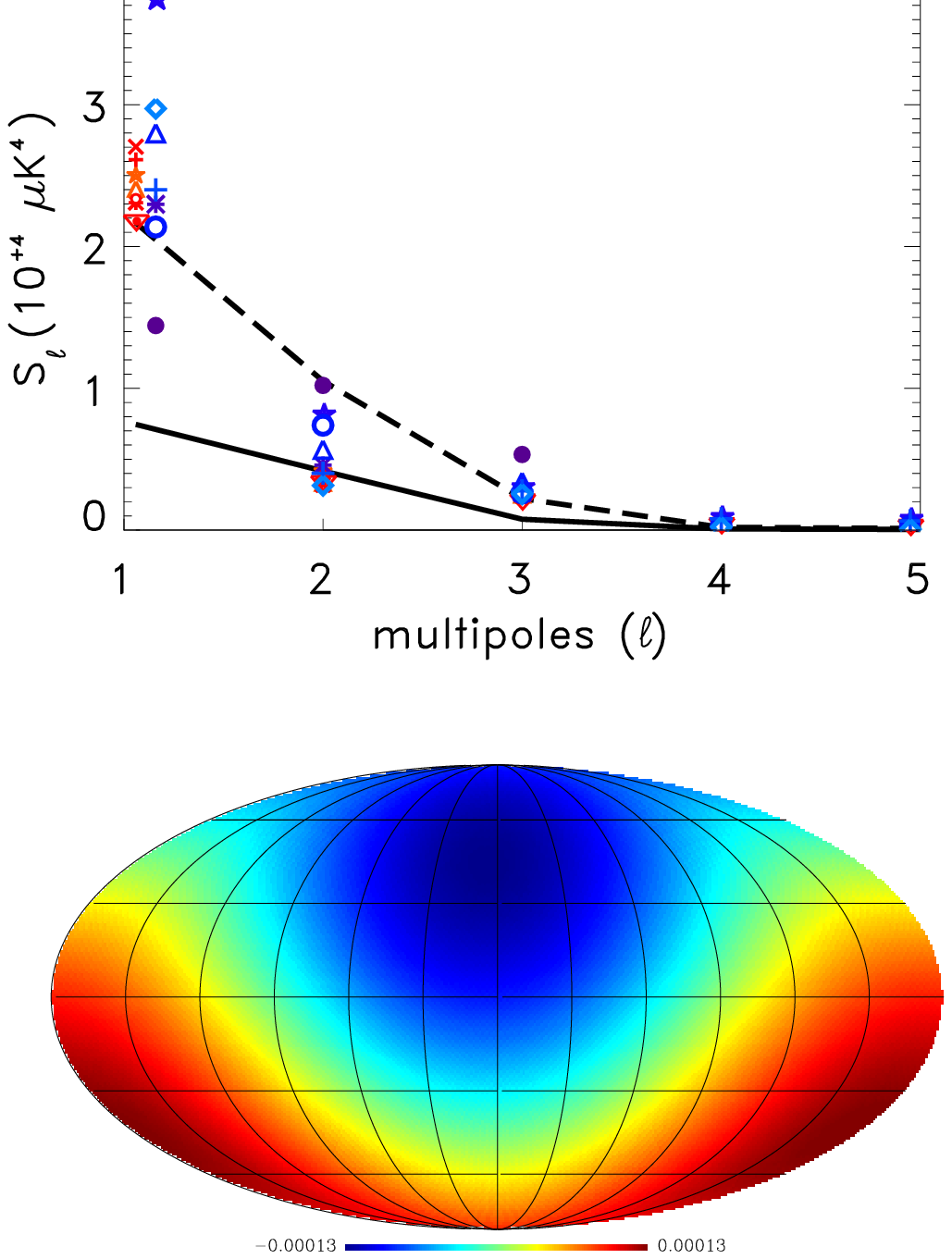}

\vspace{-9cm}
\caption{\label{fig3}
Top: Angular power spectra $\mbox{\sc S}_{\ell}$ of the 
sigma-maps-WMAP-KQ85-$\ell_{4-10}$ compared with the mean (solid line) and 
their corresponding 95\% CL (dashed line) of $1\,000$ sigma-maps MC, 
obtained from an equal number of MC-KQ85-$\ell_{4-10}$ CMB maps. 
The symbols representing data from three-year WMAP maps are plotted 
(in red) close to the vertical axis, while the symbols representing data 
from the five-year WMAP maps are plotted (in blue) slightly shifted to 
the right. The Q,  V,        W,    VW,       QVW,    OT,   PPG,  KNC,    
and ILC CMB maps are represented by the bullet, asterisk, star, triangle, 
circle, nabla, {\bf\large $\times$}, diamond, and {\bf +} symbols, 
respectively. 
The NS-asymmetry phenomenon in WMAP CMB maps is evidenced, except for the 
Q-5yr map, through the anomalously large dipole value of the 
sigma-maps-WMAP KQ85, at more than 95\% CL. 
Bottom: This is the dipole term of the sigma-maps-ILC-5yr KQ85 obtained 
with $\gamma_0 = 45^{\circ}$ (actually the $\gamma_0 = 30^{\circ}, 60^{\circ}$ 
cases are similar), where we observe that it points in the direction 
$(l, b) \simeq (190^{\circ},\, 130^{\circ})$. 
} 
\end{figure}

\begin{figure} 
\includegraphics[width=15cm,height=24cm]{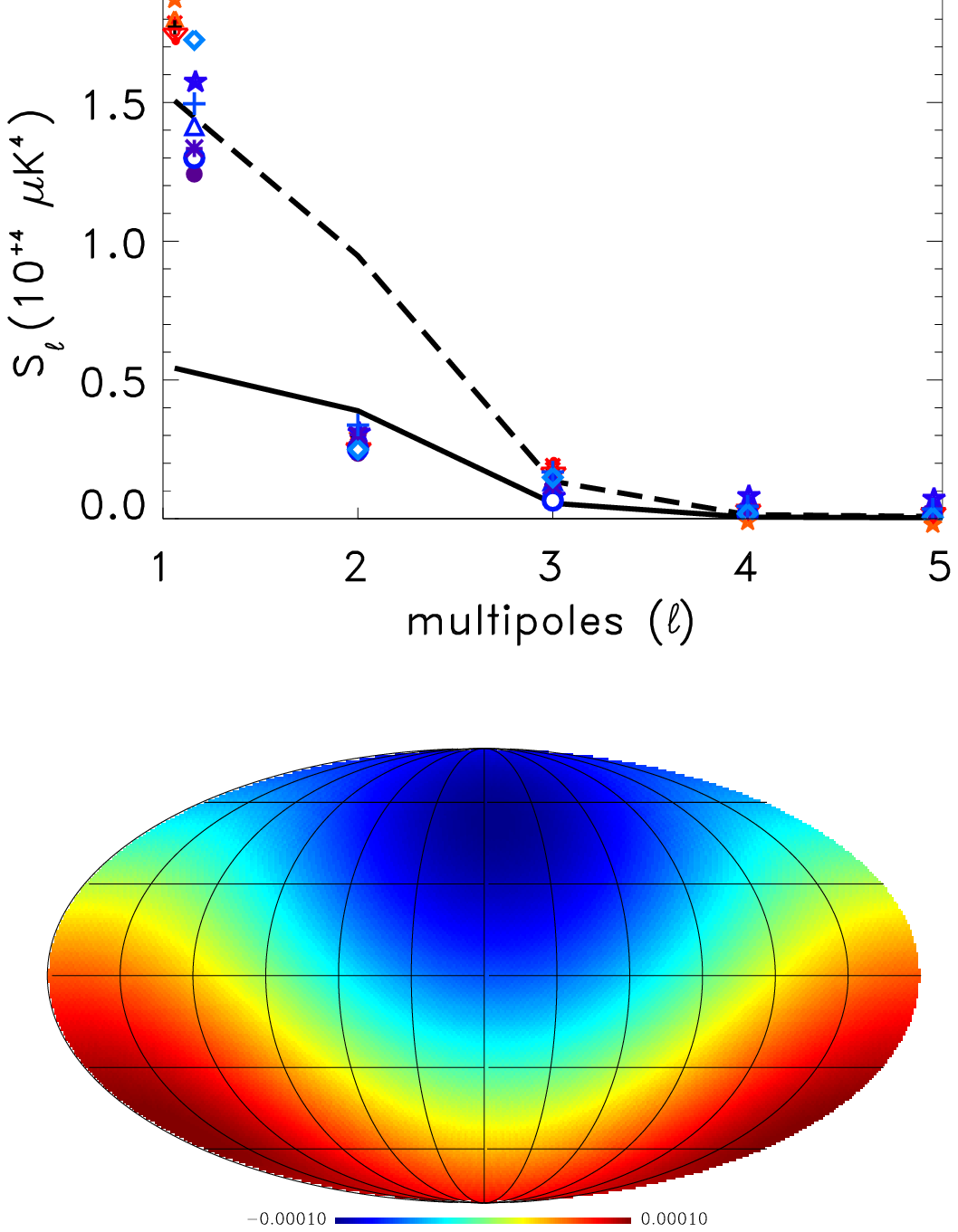}

\vspace{-9cm}
\caption{\label{fig4}
Top: Angular power spectra $\mbox{\sc S}_{\ell}$ of the 
sigma-maps-WMAP-KQ75-$\ell_{4-10}$ compared with the mean (solid line) and 
their corresponding 95\% CL (dashed line) of $1\,000$ sigma-maps MC, 
obtained from an equal number of MC-KQ75-$\ell_{4-10}$ CMB maps. 
The symbols representing data from three-year WMAP maps are plotted 
(in red) close to the vertical axis, while the symbols representing data 
from the five-year WMAP maps are plotted (in blue) slightly shifted to 
the right. The Q,  V,        W,    VW,       QVW,    OT,   PPG,  KNC,    
and ILC CMB maps are represented by the bullet, asterisk, star, triangle, 
circle, nabla, {\bf\large $\times$}, diamond, and {\bf +} symbols, 
respectively. 
We observe that the NS-asymmetry phenomenon in WMAP CMB maps is revealed 
at more than 98\% CL, in all the single, multifrequency, and ILC-type 
WMAP-3yr-Kp0 maps and also in the KNC-5yr-KQ75 foreground-cleaned map. 
However, in all the other sigma-maps-WMAP-5yr-KQ75 maps the statistical 
significance is lower, and this matter has been suitably analyzed in 
Sec.~\ref{results2}. 
Bottom: This is the dipole term of the sigma-maps-ILC-5yr KQ75 obtained with 
$\gamma_0 = 45^{\circ}$ (actually the $\gamma_0 = 30^{\circ}, 60^{\circ}$ 
cases are similar), where we observe that it points in the direction 
$(l, b) \simeq (180^{\circ},\, 130^{\circ})$. 
} 
\end{figure}

We found that the single-frequency Q-3yr, V-3yr and W-3yr WMAP maps show, 
independently of the mask (Kp2 or Kp0) used to obtain the multipoles 
$4 \leq \ell \leq 10$, the same sigma-map angular power spectra with confidence 
level $\sim 96\%$ in all the cases (see in the top panel of Fig.~\ref{fig3}, symbols 
plotted close to the vertical line). 
This result indicates, at least concerning the sigma-map analysis, that the 
single-frequency Q-3yr, V-3yr, and W-3yr maps are reasonable foreground cleaned. 
As a consequence, the multifrequency CMB maps, that is VW, QVW, and ILC-type, 
produced with these single-frequency maps behaves accordingly. 
For the WMAP-5yr-KQ85 data (see in the top panel of Fig.~\ref{fig3}, symbols plotted 
slighty shifted to the right) we found in all, but one, of these CMB maps, 
at 95\% $-$ 99\% CL, an uneven hemispherical distribution in the intensity of 
the two-point large-angle temperature correlations. 
The exceptional case refers to the Q-5yr map. 
On the contrary the single-frequency Q-5yr, V-5yr and W-5yr WMAP maps exhibit 
a disparate behavior as seen in their corresponding sigma-map angular power spectra, 
in both KQ85 and KQ75 mask cases. 
This probably suggests that these CMB maps have residual foregrounds that diminish 
but are not fully eliminated with the use of the more severe KQ75 mask. 

\begin{figure} 
\vspace{-1cm}
\includegraphics[width=15cm,height=24cm]{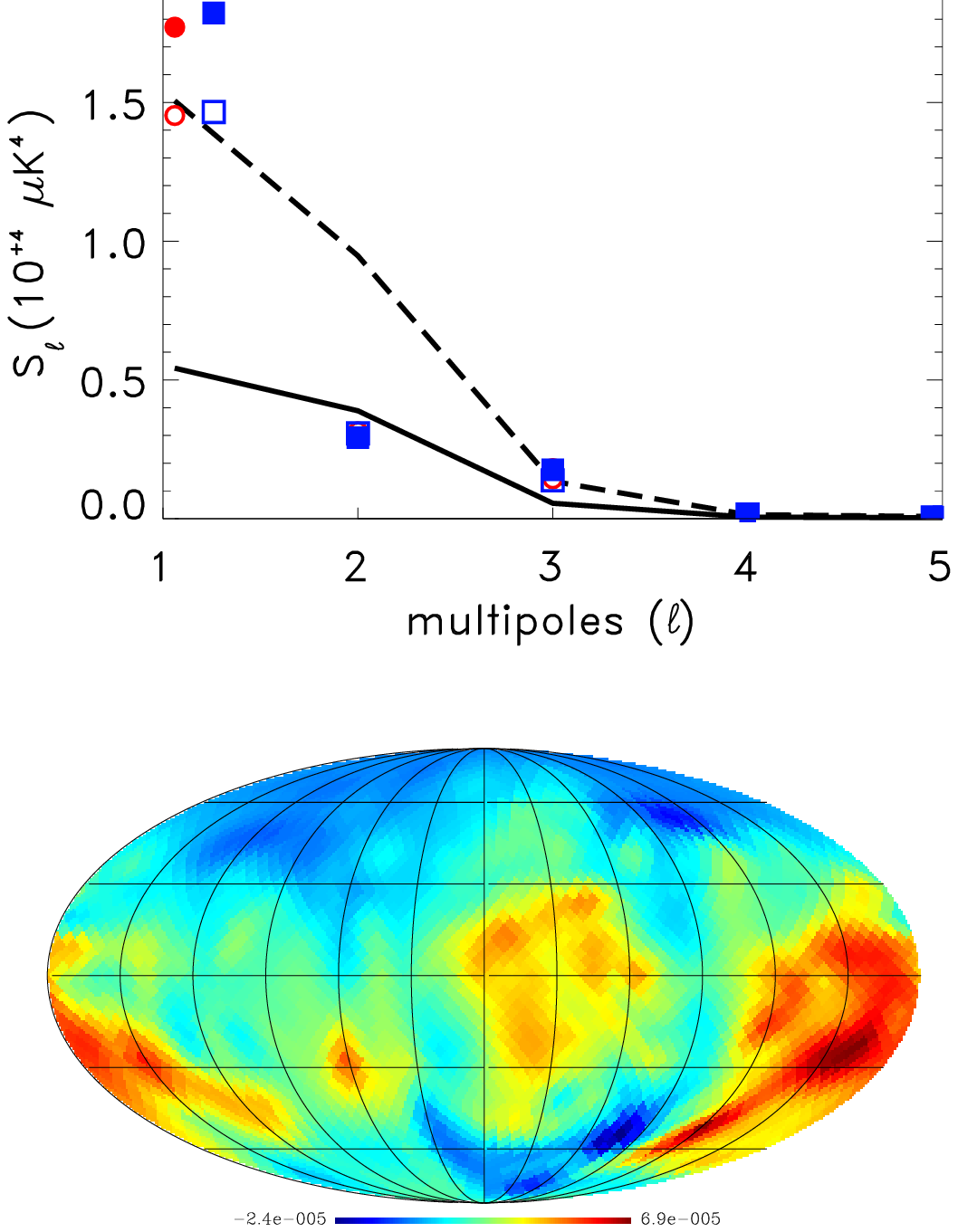}

\vspace{-9cm}
\caption{\label{fig5}
Top: Sigma-map angular power spectra of the analyze CMB maps: 
ILC-3yr-KQ75, ILC-3yr-Kp0, 
ILC-5yr-KQ75, and ILC-5yr-Kp0, represented 
as circle, filled-circle (or bullet), square, and filled-square symbols, 
respectively. 
These results show that 
$\mbox{\sc S}_{\ell}^{_{\mbox{\footnotesize ILC-3yr-KQ75}}} \simeq 
 \mbox{\sc S}_{\ell}^{_{\mbox{\footnotesize ILC-5yr-KQ75}}} $ and 
$\mbox{\sc S}_{\ell}^{_{\mbox{\footnotesize ILC-5yr-Kp0}}} \simeq 
 \mbox{\sc S}_{\ell}^{_{\mbox{\footnotesize ILC-3yr-Kp0}}} $, 
demonstrating that the main reason for the dif\/ferent statistical 
significance found in Fig.~\ref{fig4} for the sigma-map-ILC-3yr and 
sigma-map-ILC-5yr is due to the distinct cut-sky area defined by the 
KQ75 and Kp0 masks. 
In fact, the NS-asymmetry phenomenon is present, at more than 96\% CL, 
in those maps where the Kp0 mask has been applied (i.e., the 
ILC-3yr-Kp0 and ILC-5yr-Kp0 CMB maps). 
Bottom: This is the dif\/ference map: 
sigma-map-ILC-5yr-Kp0 minus sigma-map-ILC-5yr-KQ75, which clearly reveals 
through its red-intense spots that the application of dif\/ferent masks has 
a significant impact on the detection of the NS-asymmetry phenomenon: 
the KQ75 mask cuts a sky patch that turns out to be critical for the 
computation of the large-scale 2PACF in the sigma-maps. 
} 
\end{figure}

We also notice, despite that the ILC-5yr-KQ75 and the ILC-3yr-Kp0 CMB maps are fully 
correlated (Pearson's correlation coef\/ficient: 0.985180 and 0.985466, for the CMB 
data outside the KQ75 and Kp0 masks) their corresponding sigma maps are not so much. 
This fact motivated a detailed examination of the masking ef\/fects on the sigma-map 
angular power spectra. 
To estimate the inf\/luence of the dif\/ferent masks employed we calculate the 
sigma maps for the ILC-3yr CMB map obtained after applying the five-year KQ75 mask, 
and for the ILC-5yr CMB map after applying the three-year Kp0 mask. 
The angular power spectra of sigma-map-3yr-KQ75, sigma-map-3yr-Kp0, 
sigma-map-5yr-KQ75, and sigma-map-5yr-Kp0 data are shown in the top panel of 
Fig.~\ref{fig5}. 
We conclude that the KQ75 mask, as compared with Kp0, has a significant impact 
on the detection of the NS-asymmetry phenomenon decreasing its statistical 
significance. 
In fact, the KQ75 mask cuts a sky patch that turns out to be critical for the 
computation of the large-scale 2PACF in the sigma maps (as revealed by the 
dif\/ference map: sigma-map-ILC-5yr Kp0 minus sigma-map-ILC-5yr KQ75, 
plotted at the bottom panel of Fig.~\ref{fig5}). 

\begin{figure} 
\includegraphics[width=15cm,height=24cm]{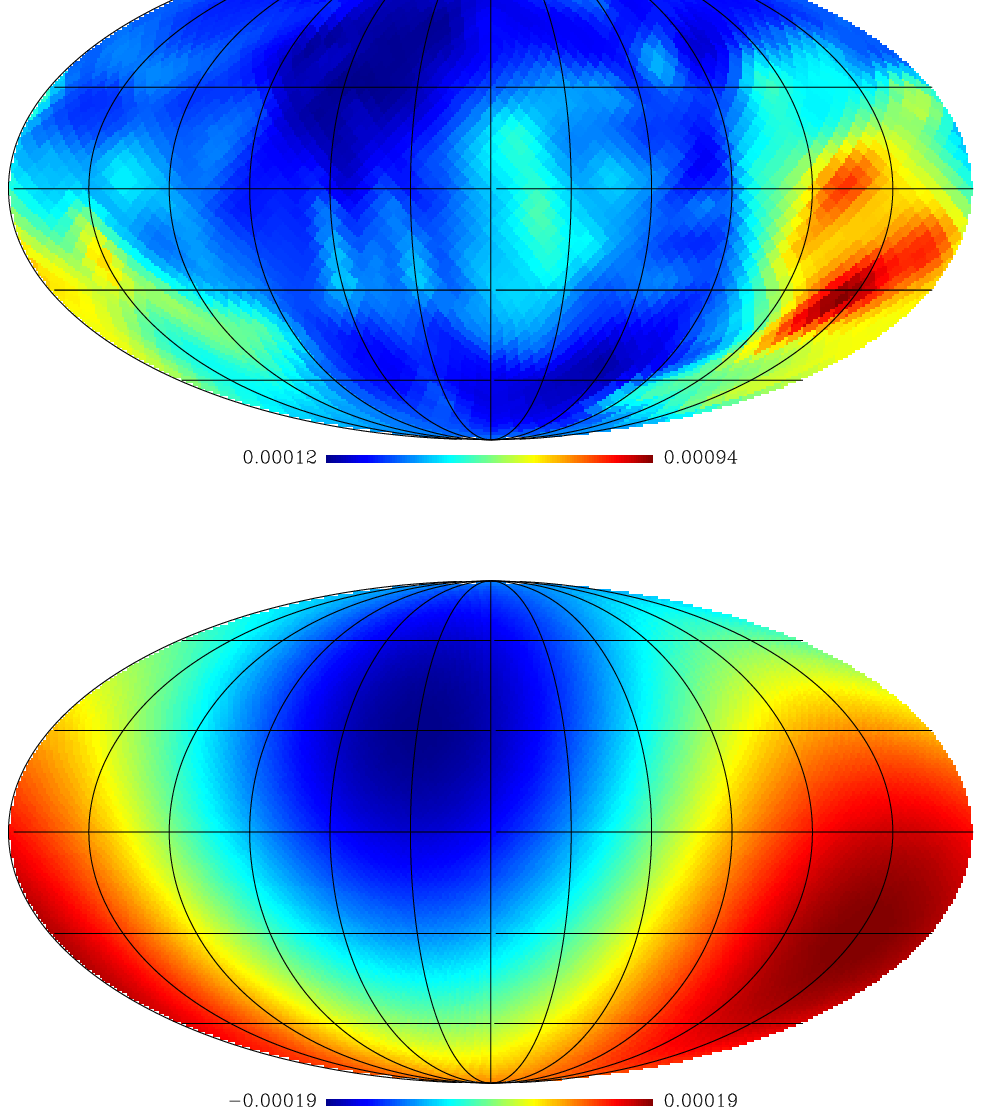}

\vspace{-10cm}
\caption{\label{fig6}
For illustration we show the sigma-map-ILC-5yr-$\ell_{4-10}$ computed from 
the WMAP-ILC-5yr full-sky CMB map (top panel) and its corresponding dipole 
term (bottom panel). 
Observe that this dipole term is pointing in the direction 
$(l, b) \simeq (220^{\circ},\, 120^{\circ})$.
} 
\end{figure}

Summarizing, 
first, our results prove that the NS-asymmetry phenomenon is present at the large 
angular scales defined by the CMB multipoles $4 \le \ell \le 10$, at a high 
significance level, in a set of single and multifrequency CMB maps from three- 
and five-year WMAP data, 
where the statistical significance depends on both, the CMB map analyzed as well 
as the cut-sky mask applied to exclude foregrounds. 
In general, we obtain that the NS-asymmetry is statistically less significant in 
five-year WMAP data with KQ75 ($\gtrsim 90\%$ CL) than it is in three-year data 
with Kp0 ($\gtrsim 96\%$ CL). 
For instance, in the ILC-5yr-KQ75 map the NS-asymmetry is observed at 92.7\% CL, 
whereas in the ILC-3yr-Kp0 map this phenomenon appears at 96.5\% CL. 
Second, these results also show that the NS-asymmetry in WMAP maps is unrelated 
to the quadrupole and octopole CMB components because they were removed from the 
CMB maps in our analysis. 
Third, due to the whole set of sensitivity tests performed our results strongly 
suggest that this hemispherical power asymmetry, is unlikely correlated 
with spurious phenomena like residual foregrounds, masking applications, or 
known systematic ef\/fects. 
Additionally, we found that the statistical significance of the NS-asymmetry 
phenomenon is af\/fected by the use of the KQ85 and KQ75 masks in  five-year 
single-frequency WMAP CMB maps, whereas it is insensitive with respect to 
application of the Kp2 and Kp0 masks in three-year single-frequency WMAP CMB maps. 
Moreover, we also examined satisfactorily the robustness of our results under 
several parameters used in the application of the sigma-map method. 
The statistical significance of the sigma-maps-WMAP angular power spectra were 
calculated by comparing them against those obtained from $1\,000$ Monte Carlo 
CMB maps produced in agreement with the WMAP $\Lambda$CDM concordance 
model~\cite{Nolta08}.
Finally, as an illustrative case, we also performed the sigma-map analysis for 
the full-sky ILC-5yr-$\ell_{4-10}$ CMB map, where we find the NS-asymmetry 
phenomenon at 95.8\% CL for the angular scales $4 \le \ell \le 10$, a result 
that is independent of the quadrupole and octopole CMB components. 
For completeness, we plot in Fig.~\ref{fig6} the sigma-map-ILC-5yr-$\ell_{4-10}$ 
(top panel) and its corresponding dipole term (bottom panel), which reveals that 
the NS-asymmetry axis for the ILC-5yr-$\ell_{4-10}$ map points along the direction 
$(l, b) \simeq (220^{\circ},\, 120^{\circ})$.

%%%%%%%%%%%%%%%%%%%%%%%%%%%%%% FIGURA 1 %%%%%%%%%%%%%%%%%%%%%%%%%%%%%%%%%%%%%%%%
%%%%%%%%%%%%%%%%%%%%%%% FIGURA 1 %%%%%%%%%%%%%%%%%%%%%%%%%%%%%%%%%%%%%%%%%%%%%

%%%%%%%%%%%%%%%%%%%%%%% FIGURA 2 %%%%%%%%%%%%%%%%%%%%%%%%%%%%%%%%%%%%%%%%%%%%%
%%%%%%%%%%%%%%%%%%%%%%% FIGURA 2 %%%%%%%%%%%%%%%%%%%%%%%%%%%%%%%%%%%%%%%%%%%%

%%%%%%%%%%%%%%%%%%%%%%% FIGURA 3 %%%%%%%%%%%%%%%%%%%%%%%%%%%%%%%%%%%%%%%%%%%%
%%%%%%%%%%%%%%%%%%%%%%% FIGURA 3 %%%%%%%%%%%%%%%%%%%%%%%%%%%%%%%%%%%%%%%%%%%%

%%%%%%%%%%%%%%%%%%%%%%% FIGURA 4 %%%%%%%%%%%%%%%%%%%%%%%%%%%%%%%%%%%%%%%%%%%%
%%%%%%%%%%%%%%%%%%%%%%%%%%% FIGURA 4 %%%%%%%%%%%%%%%%%%%%%%%%%%%%%%%%%%%%%%%%%%%

%%%%%%%%%%%%%%%%%%%%%%%%%%% FIGURA 5 %%%%%%%%%%%%%%%%%%%%%%%%%%%%%%%%%%%%%%%%%%%
%%%%%%%%%%%%%%%%%%%%%%%%%%%%% FIGURA 5 %%%%%%%%%%%%%%%%%%%%%%%%%%%%%%%%%%%%%%%%%

%%%%%%%%%%%%%%%%%%%%%%%%%%%%% FIGURA 6 %%%%%%%%%%%%%%%%%%%%%%%%%%%%%%%%%%%%%%%%%
%%%%%%%%%%%%%%%%%%%%%%%%%%% FIGURA 6 %%%%%%%%%%%%%%%%%%%%%%%%%%%%%%%%%%%%%%%%%%%%%

\section*{Acknowledgements}
\noindent
I am grateful for the use of the Legacy Archive for Microwave Background 
Data Analysis (LAMBDA). 
Some of the results in this paper have been derived using the HEALPix 
package~\cite{Gorski}. 
I acknowledge I.S. Ferreira, C.A. Wuensche, A.F.F. Teixeira, and T. Villela 
for insightful comments and suggestions. 
A.B. is supported by MCT PCI/DTI.

%%%%%%%%%%%%%%%%%%%%%%%%%%%%%%%%%%% FIM %%%%%%%%%%%%%%%%%%%%%%%%%%%%%%%%%%%%%%%%%%%%%%%%%%%%%

\begin{thebibliography}{99}
% arXiv: [astro-ph]
%% INICIO =========================== REFERENCIAS WMAP 5y ===========================
%%Hinshaw08,Gold08,Nolta08,Dunkley08,Komatsu08
\bibitem{Hinshaw08} G. Hinshaw {\it et al.}, arXiv:0803.0732.  
% 1. arXiv:0803.0732 
% "Five-Year Wilkinson Microwave Anisotropy Probe (WMAP) Observations: Data Processing, 
% Sky Maps, and Basic Results" 
% Authors: G. Hinshaw, J. L. Weiland, R. S. Hill, N. Odegard, D. Larson, C. L. Bennett, 
% J. Dunkley, B. Gold, M. R. Greason, N. Jarosik, E. Komatsu, M. R. Nolta, L. Page, 
% D. N. Spergel, E. Wollack, M. Halpern, A. Kogut, M. Limon, S. S. Meyer, G. S. Tucker, 
% E. L. Wright 

\bibitem{Gold08} B. Gold {\it et al.}, arXiv:0803.0715. 
% 2. arXiv:0803.0715 
% "Five-Year Wilkinson Microwave Anisotropy Probe (WMAP) Observations: Galactic 
% Foreground Emission" 
% Authors: B. Gold, C. L. Bennett, R. S. Hill, G. Hinshaw, N. Odegard, L. Page, 
% D. N. Spergel, J. L. Weiland, J. Dunkley, M. Halpern, N. Jarosik, A. Kogut, E. Komatsu, 
% D. Larson, S. S. Meyer, M. R. Nolta, E. Wollack, E. L. Wright 

\bibitem{Dunkley08} J. Dunkley {\it et al.}, arXiv:0803.0586. 
% 4. arXiv:0803.0586 
% "Five-Year Wilkinson Microwave Anisotropy Probe (WMAP) Observations: Likelihoods 
% and Parameters from the WMAP data" 
% Authors: J. Dunkley, E. Komatsu, M. R. Nolta, D. N. Spergel, D. Larson, G. Hinshaw, 
% L. Page, C. L. Bennett, B. Gold, N. Jarosik, J. L. Weiland, M. Halpern, R. S. Hill, 
% A. Kogut, M. Limon, S. S. Meyer, G. S. Tucker, E. Wollack, E. L. Wright 

\bibitem{Nolta08} M.R. Nolta {\it et al.}, arXiv:0803.0593. 
% 3. arXiv:0803.0593 
% "Five-Year Wilkinson Microwave Anisotropy Probe (WMAP) Observations: 
% Angular Power Spectra" 
% Authors: M. R. Nolta, J. Dunkley, R. S. Hill, G. Hinshaw, E. Komatsu, D. Larson, L. Page, 
% D. N. Spergel, C. L. Bennett, B. Gold, N. Jarosik, N. Odegard, J. L. Weiland, E. Wollack, 
% M. Halpern, A. Kogut, M. Limon, S. S. Meyer, G. S. Tucker, E. L. Wright 

\bibitem{Komatsu08} E. Komatsu {\it et al.}, arXiv:0803.0547. 
% 7. arXiv:0803.0547 
% "Five-Year Wilkinson Microwave Anisotropy Probe (WMAP) Observations: Cosmological 
% Interpretation" 
% Authors: E. Komatsu, J. Dunkley, M. R. Nolta, C. L. Bennett, B. Gold, G. Hinshaw, 
% N. Jarosik, D. Larson, M. Limon, L. Page, D. N. Spergel, M. Halpern, R. S. Hill, 
% A. Kogut, S. S. Meyer, G. S. Tucker, J. L. Weiland, E. Wollack, E. L. Wright
%%
%% FIM ================================ REFERENCIAS WMAP 5y =============================

%% INICIO ================================= REFERENCIAS WMAP 1y ============================
%% Bennett03a,Bennett03b,Hinshaw03
%% arXiv: [astro-ph]
\bibitem{Bennett03a} C.L. Bennett {\it et al.}, Astrophys. J. Suppl. Ser. {\bf 148}, 1 (2003).
% astro-ph/0302207
% First Year Wilkinson Microwave Anisotropy Probe (WMAP) Observations: 
% Preliminary Maps and Basic Results.
% Bennett, C. L., Halpern, M., Hinshaw, G., et al. 2003a, \apjs, 148, 1   
%
\bibitem{Bennett03b} C.L. Bennett {\it et al.}, Astrophys. J. Suppl. Ser. {\bf 148}, 97 (2003).
%astro-ph/0302208
%%\bibitem[ILC1y]
% First Year Wilkinson Microwave Anisotropy Probe (WMAP) Observations: Foreground Emission.
%
\bibitem{Hinshaw03} G. Hinshaw {\it et al.}, Astrophys. J. Suppl. Ser. {\bf 148}, 135 (2003).
%astro-ph/0302217
%\bibitem[coadded] <- o Coadded esta definido aqui
%First Year Wilkinson Microwave Anisotropy Probe (WMAP) Observations: 
%Angular Power Spectrum 
%G. Hinshaw, D. N. Spergel, L. Verde, ...
%%=================================== FIN REFERENCIAS WMAP 1y ===========================

%% INICIO ============================= REFERENCIAS WMAP 3y =============================
%% Bennett03a,Bennett03b,Hinshaw03
%% Hinshaw06,Jarosik06,Spergel06
%%
\bibitem{Hinshaw06} G. Hinshaw {\it et al.}, Astrophys. J. Suppl. Ser. {\bf 170}, 288 (2007).
%[ILC3y]
%astro-ph/0603451
%Three-Year Wilkinson Microwave Anisotropy Probe (WMAP) Observations: Temperature Analysis 
%G. Hinshaw, M. R. Nolta, C. L. Bennett, et al.

\bibitem{Jarosik06} N. Jarosik {\it et al.}, Astrophys. J. Suppl. Ser. {\bf 170}, 263 (2007).
%astro-ph/0603452
%Three-Year Wilkinson Microwave Anisotropy Probe (WMAP) Observations: 
%Beam Profiles, Data Processing, Radiometer Characterization and Systematic Error Limits 
%N. Jarosik, C. Barnes, M. R. Greason, 

\bibitem{Spergel06} D.N. Spergel {\it et al.}, Astrophys. J. Suppl. Ser. {\bf 170}, 377 (2007).
%astro-ph/0603449
%Wilkinson Microwave Anisotropy Probe (WMAP) Three Year Results: Implications for Cosmology 
%D. N. Spergel, R. Bean, O. Dor\'e, et al.
%
%\bibitem{Lambda}{\mbox {http://lambda.gsfc.nasa.gov/product/map/current/}}
%\bibitem{Lambda}{\mbox {http://lambda.gsfc.nasa.gov/product/map/dr1/imaps$_{-}$IDA.cfm }} 
%for the 1-year maps and {\mbox
%{http://lambda.gsfc.nasa.gov/product/map/dr2/maps$_{-}$da$_{-}$r9$_{-}$iqu$_{-}$fg$_{-}$3yr$_{-}$get.cfm}
%} for the 3-year maps
%%
%% FIM ================================ REFERENCIAS WMAP 3y ===============================


%INI-------------------------------* ANISOTROPY 1-y *---------------------------------------
%Hansen04a,Hansen04b,Bielewicz04,Bielewicz05,Eriksen04a,Eriksen04b,Eriksen05,Vielva04,
%Copi04,Copi05,Land04,Land05b,Chiang03,Vielva04,Cruz05,McEwen05,McEwen06,BVWLF
% arXiv: [astro-ph]
\bibitem{Hansen04a} F.K. Hansen, A.J. Banday, and K.M. G\'orski, 
Mon. Not. R. Astron. Soc. {\bf 354}, 641 (2004). 
%astro-ph/0404206
%Testing the cosmological principle of isotropy: local power spectrum estimates of the WMAP data, 

\bibitem{Hansen04b} F.K. Hansen, P. Cabella, D. Marinucci, and N. Vittorio, 
%\bibitem{Hansen04b} F. K. Hansen {\it et al.}, 
\apj {\bf 607}, L67 (2004).
%astro-ph/0402396 
%Asymmetries in the local curvature of the WMAP data

\bibitem{Bielewicz04} P. Bielewicz, K.M. G\'orski, and A.J. Banday, 
Mon. Not. R. Astron. Soc. {\bf 355}, 1283 (2004).
%astro-ph/0405007
%Low-order multipole maps of CMB anisotropy derived from WMAP

\bibitem{Eriksen04a} H.K. Eriksen, F.K. Hansen, A.J. Banday, K.M. G\'orski, 
and P. B. Lilje, 
\apj {\bf 605}, 14 (2004). %; Erratum {\it ibid.} {\bf 609}, 1198 (2004).
%\bibitem{Eriksen04a} H. K. Eriksen {\it et al.}, 
%astro-ph/0307507
%Asymmetries in the CMB anisotropy field 

\bibitem{Eriksen05} H.K. Eriksen, A.J. Banday, K.M. G\'orski, and P.B. Lilje, 
%\bibitem{Eriksen05} H. K. Eriksen {\it et al.}, 
\apj {\bf 622}, 58 (2005).
%astro-ph/0407271
%The N-point correlation functions of the first-year Wilkinson 
%Microwave Anisotropy Probe sky maps  

\bibitem{Land04}  K. Land and J. Magueijo, Mon. Not. R. Astron. Soc. {\bf 357}, 
994 (2005). 
%astro-ph/0405519 
%Cubic anomalies in WMAP 

\bibitem{Land05b} K. Land and J. Magueijo, \prl {\bf 95}, 071301 (2005).
%astro-ph/0502237   
%The axis of evil 

\bibitem{Copi05} C.J. Copi, D. Huterer, D.J. Schwarz, and G.D. Starkman, 
%\bibitem{Copi05} C. J. Copi {\it et al.}, 
Mon. Not. R. Astron. Soc. {\bf 367}, 79 (2006).
%astro-ph/0508047 
%On the large-angle anomalies of the microwave sky

\bibitem{BVWLF} A. Bernui, T. Villela, C.A. Wuensche, R. Leonardi, and I. Ferreira,  
%\bibitem{BVWLF} A. Bernui {\it et al.}, 
Astron. \& Astrophys. {\bf 454}, 409 (2006). 
%astro-ph/0601593
%%
%FIM-------------------------------* ANISOTROPY 1-y *-------------------------------------


%%INI-----------------------------* ANISOTROPY THREE-YEARS *------------------------------
%%arXiv: [astro-ph]
%Wiaux06,Copi06,Huterer06,Abramo06,Vielva06,Hansen06,Vielva07,Eriksen07,Land07,BMRT
\bibitem{Wiaux06} Y. Wiaux, P. Vielva, E. Mart\'{\i}nez-Gonz\'alez, and P. Vandergheynst, 
%\bibitem{Wiaux06} Y. Wiaux {\it et al.},
\prl {\bf 96}, 151303 (2006).
%astro-ph/0603367
%Global Universe anisotropy probed by the alignment of structures in the CMB

\bibitem{Copi06} C.J. Copi, D. Huterer, D.J. Schwarz, and G.D. Starkman, 
%\bibitem{Copi06} C. J. Copi {\it et al.}, 
\prd {\bf 75}, 023507 (2007).
%astro-ph/0605135 
%The Uncorrelated Universe: Statistical Anisotropy and the 
%Vanishing Angular Correlation Function in WMAP Years 1-3

\bibitem{PPG} C.-G. Park, C. Park, and J.R. Gott III, Astrophys. J. {\bf 660}, 959 (2007).
%astro-ph/0608129
%Cleaned Three-Year WMAP CMB MAP: magnitude of the quadrupole and alignment of large scale modes 

\bibitem{Huterer06} D. Huterer, New Astronomy Reviews {\bf 50}, 868 (2006).
%arXiv:astro-ph/0608318
%Mysteries on universe's largest observable scales

\bibitem{Hansen06} F.K. Hansen, A.J. Banday, H.K. Eriksen, K.M. G\'orski, and P. B. Lilje, 
Astrophys. J. {\bf 648}, 784 (2006).
% arXiv:astro-ph/0603308
%"Foreground Subtraction of Cosmic Microwave Background Maps using WI-FIT (Wavelet based hIgh 
% resolution Fitting of Internal Templates)" 

\bibitem{Vielva06} P. Vielva, Y. Wiaux, E. Mart\'{\i}nez-Gonz\'alez, and P. Vandergheynst, 
%\bibitem{Vielva06} P. Vielva {\it et al.},
New Astron. Rev. {\bf 50}, 880 (2006).
%astro-ph/0609147
%Steerable wavelet analysis of CMB structures alignment 

\bibitem{Vielva07} P. Vielva, Y. Wiaux, E. Mart\'{\i}nez-Gonz\'alez, and P. Vandergheynst,  
%\bibitem{Vielva07} P. Vielva {\it et al.}, 
Mon. Not. R. Astron. Soc. {\bf 381}, 932 (2007).
%astro-ph/0704.3736
%Alignment and signed-intensity anomalies in Wilkinson Microwave Anisotropy Probe data 

\bibitem{Eriksen07} H.K. Eriksen, A.J. Banday, K.M. G\'orski, F.K. Hansen, and P.B. Lilje,
%\bibitem{Eriksen07} H. K. Eriksen {\it et al.}, 
Astrophys. J. {\bf 660}, L81 (2007). 
%astro-ph/0701089
%Hemispherical power asymmetry in the three-year Wilkinson Microwave Anisotropy Probe sky maps 

\bibitem{Land07} K. Land and J. Magueijo, 
Mon. Not. R. Astron. Soc. {\bf 378}, 153 (2007). 
%astro-ph/0611518
%The Axis of Evil revisited 

\bibitem{Samal} P.K. Samal, R. Saha, P. Jain, and J.P. Ralston, 
Mon. Not. R. Astron. Soc. {\bf 385}, 1718 (2008). 
%arXiv:0708.2816 [astro-ph]. 
%"Testing Isotropy of Cosmic Microwave Background Radiation" %(2007).

\bibitem{BMRT} A. Bernui, B. Mota, M.J. Rebou\c{c}as, and R. Tavakol, 
Astron. \& Astrophys. {\bf 464}, 479 (2007).
%\bibitem{BMRT} A. Bernui {\it et al.}, 
%astro-ph/0511666  
%Mapping the large-scale anisotropy in the WMAP data

\bibitem{BMRT07} A. Bernui, B. Mota, M.J. Rebou\c{c}as, and R. Tavakol, 
%\bibitem{BMRT07} A. Bernui {\it et al.},
Int. Journal of Mod. Phys. D {\bf 16}, 411 (2007).
%astro-ph/0706.0575
%A Note on the Large-Angle Anisotropies in the Wmap Cut-Sky Maps 

\bibitem{Bunn08} E.F. Bunn and A. Bourdon, arXiv:0808.0341. 
%"Contamination cannot explain the lack of large-scale power in the cosmic 
% microwave background radiation" 
%arXiv:0808.0341 [astro-ph]

\bibitem{Monteserin} C. Monteser\'{\i}n, R.B. Barreiro, P. Vielva, E. Mart\'{\i}nez-Gonz\'alez, 
M.P. Hobson, and A.N. Lasenby, Mon. Not. R. Astron. Soc. {\bf 387}, 209 (2008). 
%"A low cosmic microwave background variance in the Wilkinson Microwave Anisotropy Probe data" 

\bibitem{Lew} B. Lew, arXiv:0808.2867.
%%
%FIN-------------------------------* ANISOTROPY THREE-YEARS *-------------------------------

%%*******************************? punto de vista isotropico ?************************%%
%% arXiv: [astro-ph]
%Hajian03,Hajian04,Souradeep04,Souradeep05,Souradeep06,Hajian06
%Statistical isotropy of CMB anisotropy from WMAP  %->  astro-ph/0308002
\bibitem{Hajian03} A. Hajian and T. Souradeep, \apj {\bf 597}, L5 (2003).
%Measuring Statistical isotropy of the CMB anisotropy
%astro-ph/0308001 

\bibitem{Hajian04} A. Hajian, T. Souradeep, and N. Cornish, \apj {\bf 618}, L63 (2005).
%Statistical Isotropy of the Wilkinson Microwave Anisotropy Probe
%Data: A Bipolar Power Spectrum Analysis, astro-ph/0406354

\bibitem{Souradeep04} T. Souradeep and A. Hajian, Pramana {\bf 62}, 793 (2004).

\bibitem{Souradeep05} T. Souradeep and A. Hajian, arXiv:0502248. 
%astro-ph/0502248
%Statistical isotropy of CMB anisotropy from WMAP 

\bibitem{Souradeep06} T. Souradeep, A. Hajian, and S. Basak, 
New Astronomy Reviews {\bf 50}, 889 (2006).
%astro-ph/0607577
%Measuring statistical isotropy of CMB anisotropy 

\bibitem{Hajian06} A. Hajian and T. Souradeep, \prd {\bf 74}, 123521 (2006).
%astro-ph/0607153
%Testing global isotropy of three-year Wilkinson Microwave Anisotropy Probe (WMAP) data: 
%Temperature analysis 
%%******************************  punto de vista isotropico  **************************%%
% arXiv: [astro-ph]

\bibitem{TOH} M. Tegmark, A. de Oliveira-Costa, and A.J.S. Hamilton, 
\prd {\bf 68}, 123523 (2003).
%astro-ph/0302496
%A high resolution foreground cleaned CMB map from WMAP.

\bibitem{OTZH} A. de Oliveira-Costa, M. Tegmark, M. Zaldarriaga, and A. Hamilton, 
\prd {\bf 69}, 063516 (2004).  
%\bibitem{OTZH} A. de Oliveira-Costa {\it et al.}, 
%astro-ph/0307282

\bibitem{Weeks04} J.R. Weeks, arXiv:0412231. 
%astro-ph/0412231
%Maxwell's Multipole Vectors and the CMB

\bibitem{Abramo06} L.R. Abramo, A. Bernui, I.S. Ferreira, T. Villela, and C.A. Wuensche, 
%\bibitem{Abramo06} L. R. Abramo {\it et al.},
\prd {\bf 74}, 063506 (2006).
%astro-ph/0604346
%Alignment tests for low CMB multipoles

%%%%%%%%%%%%%%%%%%%%%%%%%%%%%%%%%%%%%%%%%%%%%%%%%%%%%%%%%%%%%%%%%%%%%%%%%%%%%%%%%%%%%%%%%%
%Chiang03,Vielva04,Copi04,Cruz05,Cruz07,McEwen06,BTV07
\bibitem{Chiang03} L.-Y. Chiang, P.D. Naselsky, O.V. Verkhodanov, and M.J. Way, 
%\bibitem{Chiang03} L.-Y. Chiang {\it et al.}, 
\apj {\bf 590}, L65 (2003). 
% astro-ph/0303643.
% Non-Gaussianity of the derived maps from the first-year WMAP data 

\bibitem{Vielva04} P. Vielva, E. Mart\'{\i}nez-Gonz\'alez, R.B. Barreiro, 
J.L. Sanz, and L. Cay\'on, \apj {\bf 609}, 22 (2004). 
% primera referencia al COLD-SPOT
% \bibitem{Vielva04} P. Vielva {\it et al.}, 
% astro-ph/0310273.
% Detection of non-Gaussianity in the WMAP 1-year data using spherical wavelets

\bibitem{Copi04} C.J. Copi, D. Huterer, and G.D. Starkman, \prd {\bf 70}, 
043515 (2004).  
%astro-ph/0310511
%Multipole Vectors--a new representation of the CMB sky and evidence for 
%statistical anisotropy or non-Gaussianity at 2<=l<=8 

\bibitem{Cruz05} M. Cruz, E. Mart\'{\i}nez-Gonz\'alez, P. Vielva, and L. Cay\'on, 
%\bibitem{Cruz05} M. Cruz {\it et al.}, 
Mon. Not. R. Astron. Soc. {\bf 356}, 29 (2005).
%astro-ph/0405341
%Detection of a non-Gaussian spot in WMAP 

\bibitem{Cruz07} M. Cruz, L. Cay\'on, E. Mart\'{\i}nez-Gonz\'alez, P. Vielva, and J. Jin, 
%\bibitem{Cruz07} M. Cruz {\it et al.}, 
\apj {\bf 655}, 11 (2007).
%astro-ph/0603859
%The Non-Gaussian Cold Spot in the 3 Year Wilkinson Microwave Anisotropy Probe Data 

\bibitem{McEwen06} J.D. McEwen, M.P. Hobson, A.N. Lasenby, and D.J. Mortlock,  
%\bibitem{McEwen06} J. D. McEwen {\it et al.}, 
Mon. Not. R. Astron. Soc. {\bf 371}, L50 (2006).
%astro-ph/0604305
%A high-significance detection of non-Gaussianity in the WMAP 3-year data using directional 
%spherical wavelets 

\bibitem{BTV07} A. Bernui, C. Tsallis, and T. Villela, 
Europhys. Let., {\bf 78}, 19001 (2007). 
%astro-ph/0703708
%Deviation from Gaussianity in the cosmic microwave background temperature fluctuations 
%
% \bibitem{BTV05) Bernui, A., Tsallis, C., and Villela, T. Phys. Let. A, {\bf ?}, ? (2006). 
% astro-ph/0512267

\bibitem{Martinez08} E. Mart\'{\i}nez-Gonz\'alez, arXiv:0805.4157.
% arXiv:0805.4157v1 [astro-ph] 
%"Gaussianity"

\bibitem{McEwen08} J.D. McEwen, M.P. Hobson, A.N. Lasenby, and D.J. Mortlock, 
Mon. Not. R. Astron. Soc. {\bf 388}, 659 (2008). 
%arXiv:0803.2157 [astro-ph]

%arXiv:0803.2157v2 [astro-ph] 
%"A high-significance detection of non-Gaussianity in the WMAP 5-year data using 
% directional spherical wavelets"

% arXiv: [astro-ph]
%%FIN-----------------------------* ANISOTROPY 1-Y *------------------------------------

\bibitem{Aurich05} R. Aurich, S. Lustig, and F. Steiner, 
Class. Quantum Grav. {\bf 22}, 3443 (2005). 
%astro-ph/0504656
%CMB Anisotropy of Spherical Spaces 

\bibitem{Land05a} K. Land and J. Magueijo, Mon. Not. R. Astron. Soc. {\bf 367}, 
1714 (2006).
%astro-ph/0509752
%Template fitting and the large-angle CMB anomalies 

\bibitem{Jaffe06} T.R. Jaffe, A.J. Banday, H.K. Eriksen, K.M. G\'orski, 
and F. K. Hansen, 
Astron. \& Astrophys. {\bf 460}, 393 (2006). 
%\bibitem{Jaffe06} T. R. Jaffe {\it et al.}, 
%astro-ph/0606046
%Bianchi Type VII_h Models and the WMAP 3-year Data 

\bibitem{Hipolito} W.S. Hip\'olito-Ricaldi and G.I. Gomero, \prd {\bf 72}, 
103008 (2005).
%astro-ph/0507238 

\bibitem{Cresswell} J.G. Cresswell, A.R. Liddle, P. Mukherjee, and A. Riazuelo, 
%\bibitem{Cresswell} J. G. Cresswell {\it et al.},
\prd {\bf 73}, 041302(R) (2006).
%astro-ph/0512017
%Cosmic microwave background multipole alignments in slab topologies 

\bibitem{Ghosh} T. Ghosh, A. Hajian, and T. Souradeep, \prd {\bf 75}, 083007 (2007).
%astro-ph/0604279
%Unveiling hidden patterns in CMB anisotropy maps 

\bibitem{Gordon05} C. Gordon, W. Hu, D. Huterer, and T. Crawford, 
%\bibitem{Gordon05} C. Gordon {\it et al.}, 
\prd {\bf 72}, 103002 (2005). 
%astro-ph/0509301
%Spontaneous Isotropy Breaking: A Mechanism for CMB Multipole Alignments 

\bibitem{Contaldi07} A.E. G\"umr\"uk\c{c}\"uo\u{g}lu, C.R. Contaldi, and M. Peloso, 
J. Cosmol. Astropart. Phys. {\bf 11}, 005 (2007). 
%arXiv:0707.4179. 
%Inflationary perturbations in anisotropic backgrounds and their imprint on the CMB 

\bibitem{Pullen} A.R. Pullen and M. Kamionkowski, 
\prd {\bf 76}, 103529 (2007). 
%arXiv:0709.1144.
%astro-ph/0709.1144
%Cosmic Microwave Background Statistics for a Direction-Dependent Primordial Power Spectrum 

\bibitem{Ackerman} L. Ackerman, S.M. Carroll, and M.B. Wise, 
\prd {\bf 75}, 083502 (2007). 
%astro-ph/0701357
%Imprints of a primordial preferred direction on the microwave background 

\bibitem{Dutta} J.F. Donoghue, K. Dutta, and A. Ross, arXiv:0703455. 
% "Non-isotropy in the CMB power spectrum in single field inflation"
% Contaldi et al. [1] have suggested that an initial period of kinetic energy 
% domination in single field inflation may explain the lack of CMB power at large 
% angular scales. We note that in this situation it is natural that there also be a 
% spatial gradient in the initial value of the inflaton field, and that this can 
% provide a spatial asymmetry in the observed CMB power spectrum, manifest at low 
% multipoles. We investigate the nature of this asymmetry and comment on its relation 
% to possible anomalies at low multipoles. 

\bibitem{Dvorkin} C. Dvorkin, H.V. Peiris, and W. Hu, 
\prd {\bf 77}, 063008 (2008). 
%arXiv:0711.2321
%"Testable polarization predictions for models of CMB isotropy anomalies" 

\bibitem{Demianski} M. Demia\'{n}ski and A.G. Doroshkevich, \prd {\bf 75}, 
123517 (2007).
%preprint (arXiv:0702381 [astro-ph])
%Possible extensions of the standard cosmological model: anisotropy, rotation, 
%and magnetic field 

\bibitem{Kahniashvili1} T. Kahniashvili, Y. Maravin, and A. Kosowsky, 
arXiv:0806.1876. 
%"Primordial Magnetic Field Limits from WMAP Five-Year Data" 
%arXiv:0806.1876 

\bibitem{Kahniashvili2} T. Kahniashvili, G. Lavrelashvili, and B. Ratra, 
arXiv:0807.4239. 
%CMB Temperature Anisotropy from Broken Spatial Isotropy due to an Homogeneous 
%Cosmological Magnetic Field 
%arXiv:0807.4239 

\bibitem{Pereira} T.S. Pereira, C. Pitrou, J.-P. Uzan, JCAP {\bf 09}, 
006 (2007).

% \bibitem{Campanelli1} L. Campanelli, P. Cea, and L. Tedesco,  
% \prl {\bf 97} 131302, (2006); Erratum-ibid. {\bf 97} 209903 (2006). 
%"Ellipsoidal Universe Can Solve The CMB Quadrupole Problem"
%astro-ph/0606266 

\bibitem{Campanelli2} L. Campanelli, P. Cea, and L. Tedesco, 
\prd {\bf 76}, 063007 (2007). 
%"Cosmic Microwave Background Quadrupole and Ellipsoidal Universe"
%arXiv:0706.3802 [astro-ph] 

\bibitem{Morales} J.A. Morales and D. S\'aez, 
Astrophys. J. {\bf 678}, 583 (2008). 
%arXiv:0802.1042 [astro-ph]. 
%"Large scale vector modes and the first CMB temperature multipoles"

\bibitem{BH} A. Bernui and W.S. Hip\'olito-Ricaldi, arXiv:0807.1076. 
%"Can a primordial magnetic field originate large-scale anomalies in WMAP data?" 

\bibitem{Rakic07} A. Raki\'c and D.J. Schwarz, \prd {\bf 75}, 103002 (2007).
%astro-ph/0703266
%Correlating anomalies of the microwave sky: The Good, the Evil and the Axis

%%>>>>>>>>>>>>>>>>>>>>>>>>>>>>>>>  FOREGROUNDS  <<<<<<<<<<<<<<<<<<<<<<<<<<<<<<<<<<<<
%%
%Eriksen04b,WW,OT,Cruz06,ASW,Chiang07,Lopez07
\bibitem{Eriksen04b} H.K. Eriksen, A.J. Banday, K.M. G\'orski, and P.B. Lilje,
%\bibitem{Eriksen04b} H. K. Eriksen {\it et al.},
\apj {\bf 612}, 633 (2004). 
%astro-ph/0403098.
%On foreground removal from the WMAP data by an Internal Linear 
%Combination Method: Limitations and Implications

\bibitem{WW} T. Wibig and A.W. Wolfendale, Mon. Not. R. Astron. Soc. {\bf 360}, 236 (2005).
%astro-ph/0409397
%Foreground contributions to the Cosmic Microwave Background 

\bibitem{OT} A. de Oliveira-Costa and M. Tegmark, \prd {\bf 74}, 023005 (2006). 
%astro-ph/0603369
%CMB multipole measurements in the presence of foregrounds 

\bibitem{Cruz06} M. Cruz, M. Tucci, E. Mart\'{\i}nez-Gonz\'alez, and P. Vielva, 
%\bibitem{Cruz06} M. Cruz {\it et al.},
Mon. Not. R. Astron. Soc. {\bf 369}, 57 (2006).
%astro-ph/0601427
%The non-Gaussian cold spot in WMAP: significance, morphology and foreground contribution 

\bibitem{ASW} L.R. Abramo, L. Sodr\'e Jr., and C.A. Wuensche, 
\prd {\bf 74}, 083515 (2006).
%astro-ph/0605269
%Anomalies in the low CMB multipoles and extended foregrounds 

\bibitem{Chiang07} L.-Y. Chiang, P. Coles, P.D. Naselsky, and P. Olesen, 
%\bibitem{Chiang07} L.-Y. Chiang {\it et al.},
JCAP {\bf 1}, 21 (2007). 
%astro-ph/0608421 
%The one-dimensional Fourier representation and large angular scale foreground 
%contamination in the three-year Wilkinson Microwave Anisotropy Probe data 

\bibitem{Lopez07} M. L\'opez-Corredoira, 
J. Astrophys. Astron. {\bf 28}, 101 (2007).
%arXiv:0708.4133 [astro-ph]. 
%Some doubts on the validity of the foreground Galactic contribution subtraction from 
%microwave anisotropies 
%%
%%>>>>>>>>>>>>>>>>>>>>>>>>>>> FIN dos >>>>> FOREGROUNDS <<<<<<<<<<<<<<<<<<<<<<<<<<<<<<<

\bibitem{Helling06} R.C. Helling, P. Schupp, and T. Tesileanu, \prd {\bf 74}, 063004 (2006).
%astro-ph/0603594) 
%CMB statistical anisotropy, multipole vectors, and the influence of the dipole

\bibitem{Bunn07} E.F. Bunn, \prd {\bf 75}, 083517 (2007).
%astro-ph/0607312
%Systematic Errors in Cosmic Microwave Background Interferometry 

\bibitem{Naselsky07} P.D. Naselsky, O.V. Verkhodanov, and M.T.B. Nielsen, 
arXiv:0707.1484.
%"Instability of reconstruction of the low CMB multipoles"

\bibitem{Schwarz04} D.J. Schwarz, G.D. Starkman, D. Huterer, and C.J. Copi,  
%\bibitem{Schwarz04} D. J. Schwarz {\it et al.},
\prl {\bf 93}, 221301 (2004). 
%astro-ph/0403353
%Is the low-$\ell$ microwave background cosmic? 

%Bielewicz05
\bibitem{Bielewicz05} P. Bielewicz {\it et al.}, Astrophys. J. {\bf 635}, 750 (2005).
%P. Bielewicz, H.K. Eriksen, A.J. Banday, K.M. G\'orski, and P.B. Lilje
% astro-ph/0507186
% Multipole vector anomalies in the first-year WMAP data: a cut-sky analysis

%\bibitem[G\'orski et al.(1998)]{Gorski} K. M. G\'orski, E. Hivon, \& 
%B. D. Wandelt, astro-ph/9812350, also at 
%$\langle$ http://www.eso.org/science/healpix/ $\rangle$ (1998). 
\bibitem{Gorski} K.M. G\'orski {\it et al.}, \apj {\bf 622}, 759 (2005).
%astro-ph/0409513
%K. M. G\'orski, E. Hivon, A. J. Banday, B. D. Wandelt, F. K. Hansen, M. Reinecke, M. Bartelmann
%"HEALPix: A Framework for High-Resolution Discretization and Fast Analysis of Data 
%Distributed on the Sphere" 
%http://healpix.jpl.nasa.gov/

\bibitem{Padmanabhan} T. Padmanabhan, 
{\it Structure formation in the universe}, (Cambridge Univ. Press, Cambridge, 1993).

%%%%%%%%%%%%%%%%%%%  outros cleaned-maps  %%%%%%%%%%%%%%%%%%%%%%%%%%%%%%%%%%%%%%%%%%
\bibitem{KNC} J. Kim, P. Naselsky, and P.R. Christensen, \prd {\bf 77}, 103002 (2008). 
%arXiv:0803.1394 [astro-ph].
% astro-ph/0803.1394
% "CMB map derived from the WMAP data through Harmonic Internal Linear Combination" 

% \bibitem{OT} A. de Oliveira-Costa and M. Tegmark, \prd \textbf{74}, 023005 (2006). 
% astro-ph/0603369
% "CMB multipole measurements in the presence of foregrounds" 

%\bibitem{PPG} C.-G. Park, C. Park, and J.R. Gott III, 
%Astrophys. J. \textbf{660}, 959 (2007).
%astro-ph/0608129
%Cleaned Three-Year WMAP CMB MAP: magnitude of the quadrupole and alignment of large scale modes 
%%%%%%%%%%%%  outros cleaned-maps  %%%%%%%%%%%%%%%%%%%%%%%%%%%%

\end{thebibliography}
\end{document}